\documentclass[%
 preprint,
 superscriptaddress,
 amsmath,amssymb,
 aps,
 showkeys,
 titlepage,
 endfloats*.   %
]{revtex4-2}

\usepackage[utf8]{inputenc}
\usepackage[english]{babel}

\usepackage{graphicx}%
\usepackage{dcolumn}%
\usepackage{bm}%

\usepackage[colorlinks = true,
            linkcolor = blue,
            urlcolor  = blue,
            citecolor = red,
            anchorcolor = blue]{hyperref}

\begin{document}

\title{Magnetic Polarons Enable Exceptional Magnetocaloric Response} 

\author{Joshua Ancheta}
\thanks{These authors contributed equally.}
\affiliation{Department of Mechanical Engineering, University of California, Santa Barbara, CA 93106, USA}

\author{Caeli Benyacko}
\thanks{These authors contributed equally.}
\affiliation{Materials Department, University of California, Santa Barbara, CA 93106, USA}

\author{Fanghao Zhang}
\thanks{These authors contributed equally.}
\affiliation{Department of Mechanical Engineering, University of California, Santa Barbara, CA 93106, USA}

\author{Stephen D. Wilson}
\email{stephendwilson@ucsb.edu}
\affiliation{Materials Department, University of California, Santa Barbara, CA 93106, USA}

\author{Bolin Liao}
\email{bliao@ucsb.edu} \affiliation{Department of Mechanical Engineering, University of California, Santa Barbara, CA 93106, USA}

\date{\today}

\begin{abstract}
Magnetocaloric materials are typically limited by a trade-off between magnetic entropy and field responsiveness. Here we show that magnetic polarons provide an intermediate regime that mitigates this constraint and enables an exceptional magnetocaloric response. Using EuB$_6$ as a model system, we combine thermodynamic and magnetic measurements to demonstrate that nanoscale ferromagnetic clusters emerging near the Curie temperature strongly enhance the field-induced entropy collapse. These clusters possess large effective moments that respond efficiently to applied fields while retaining substantial entropy due to their small size and dynamic fluctuations. As a result, EuB$_6$ exhibits a giant cryogenic magnetocaloric response, with both large isothermal entropy change and adiabatic temperature change in the technologically important 10-40 K range. Our results identify magnetic polarons as an underexplored route for optimizing magnetocaloric performance and establish an intermediate magnetic length scale as a design principle for high-performance cryogenic cooling materials.
\end{abstract}

\maketitle

\section{Introduction}
Efficient refrigeration in the 10-40 K temperature range is increasingly important for technologies spanning hydrogen liquefaction, superconducting systems, cryogenic sensors and emerging quantum platforms~\cite{Ghafri2022hydrogen,radenbaugh2004refrigeration,farrah2019far}. Magnetocaloric refrigeration offers an appealing solid-state alternative to conventional gas-compression approaches~\cite{numazawa2014magnetic,ansarinasab2023conceptual,feng2020modeling} because it can, in principle, provide high efficiency, compact form factors and vibration-free operation~\cite{kitanovski2020energy}. Yet the development of practical cryogenic magnetocaloric materials remains challenging~\cite{pecharsky1999magnetocaloric}. While many magnetic materials exhibit sizable entropy changes near their ordering temperatures, only a limited number combine large isothermal entropy change and large adiabatic temperature change under moderate magnetic fields, particularly in the technologically important 10-40 K range~\cite{kumar2024exploring}. Recent efforts have increasingly focused on materials exhibiting first-order magnetic transitions\cite{guillou2014taming,pecharsky1997giant}, which can produce giant entropy changes through coupled magnetic and structural transformations. However, these materials often suffer from thermal and magnetic hysteresis, which degrades reversibility and efficiency and limits their practical applicability.

An essential obstacle is that the two ingredients required for a large magnetocaloric response are often fundamentally in competition. On the one hand, maximizing the zero-field magnetic entropy favors systems composed of many weakly correlated atomic moments. On the other hand, strong responsiveness to an applied magnetic field favors larger collective magnetic entities that are more easily aligned. These two limits are exemplified by conventional paramagnets~\cite{pecharsky1999magnetocaloric} and superparamagnetic particles~\cite{mcmichael1992magnetocaloric}, respectively. Paramagnets possess a large entropy reservoir but respond only weakly to field because each moment is atomic in scale. In contrast, large correlated clusters or superparamagnetic particles exhibit strong field susceptibility because they behave as macrospins, but they sacrifice entropy density because many spins are already locked into a single correlated object. This competition suggests that an optimal magnetocaloric material may not reside in either extreme, but rather in an intermediate regime where substantial entropy and strong field tunability coexist.

\begin{figure}[!htb]
\includegraphics[width=\textwidth]{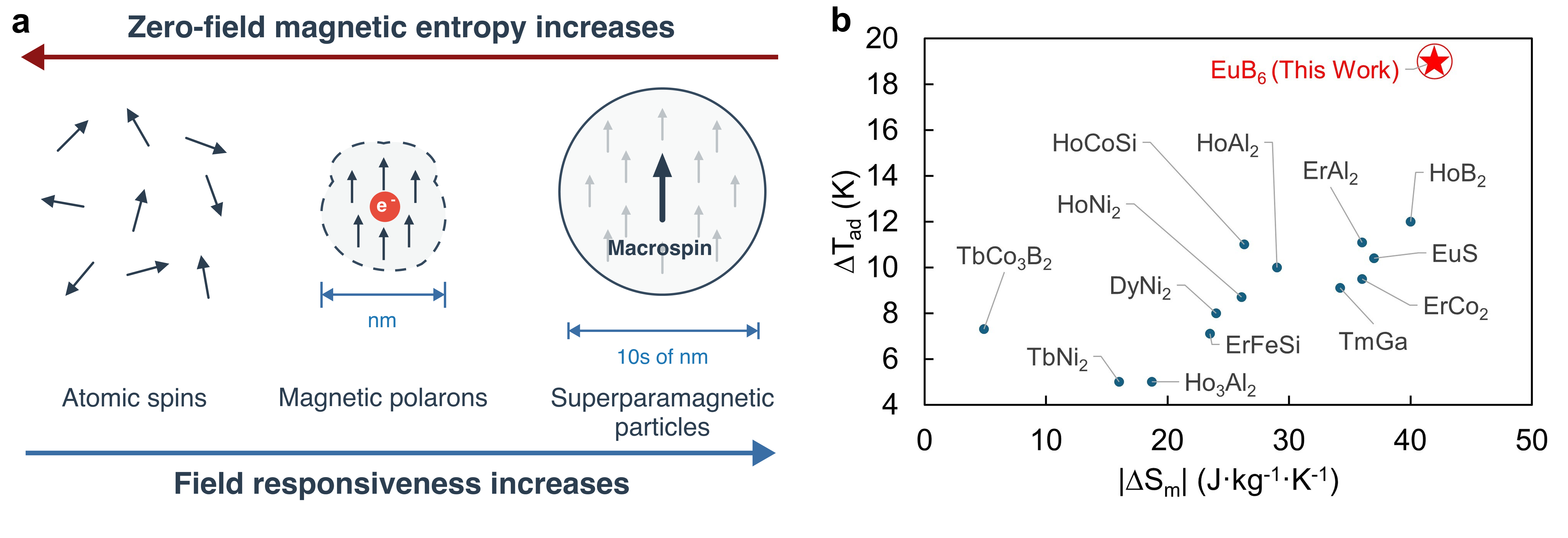}
\caption{\textbf{Magnetic polarons as a ``sweet spot'' for enhanced magnetocaloric performance.} \textbf{a}, Conceptual illustration of magnetic degrees of freedom across different length scales. Atomic spins (left) possess large configurational entropy but respond weakly to external magnetic fields. Superparamagnetic particles (right) exhibit strong field responsiveness due to large collective moments (``macrospins''), but their entropy density is reduced. Magnetic polarons (middle), consisting of nanoscale clusters of correlated spins, provide an intermediate regime in which entropy and field tunability are balanced, offering a favorable combination for magnetocaloric applications. \textbf{b}, Benchmark comparison of magnetocaloric performance for materials with Curie temperatures in the range of 10-40 K. EuB$_6$ (red star) exhibits simultaneously large 
isothermal entropy change $|\Delta S_M|$ and 
adiabatic temperature change $\Delta T_{ad}$, outperforming previously reported materials in this temperature window, highlighting the effectiveness of magnetic-polaron-mediated entropy tuning. Data are taken from the following literature \cite{kumar2024exploring,castro2020machine,YAMAMOTO2003Entropy,IbarraGaytan2013Entropy,DONG2015Entropy,CWIK2017Entropy,ZHANG2012Tad,LI2011Tad}.} 
\label{fig:fig1}
\end{figure}

Here we propose that magnetic polarons naturally realize such an optimal intermediate state. Magnetic polarons are nanoscale ferromagnetic droplets embedded in a fluctuating paramagnetic background, typically stabilized by carrier-mediated exchange interactions~\cite{mauger1983magnetic,teresa1997evidence}. Unlike isolated atomic spins, they possess collective moments that can respond strongly to external magnetic fields. Unlike rigid macroscopic magnetic clusters, however, they retain substantial magnetic entropy because they are smaller in size (on the order of nanometers), thermally fluctuating, and strongly tunable by temperature and field. As schematically illustrated in Fig.~\ref{fig:fig1}a, magnetic polarons can therefore occupy a ``sweet spot'' between entropy reservoir and field responsiveness. Despite the long-standing interest in magnetic polarons in correlated magnetic semiconductors and related systems~
\cite{koepsell2019imaging,maksimov2000magnetic,durst2002bound}, their implications for the magnetocaloric effect have remained largely unexplored.

EuB$_6$ provides an ideal platform to test this idea. It is a canonical magnetic-polaron system in which carrier-mediated ferromagnetic droplets emerge above the Curie temperature and evolve into long-range ferromagnetic order upon cooling~\cite{snow2001magnetic,brooks2004magnetic,zhang2009nonlinear,beaudin2026observation}. The material has long attracted attention because of its unusual magnetotransport~\cite{guy1980charge}, percolative magnetic transition~\cite{henggeler1998magnetic} and intertwined charge-spin dynamics~\cite{sullow2000metallization}, which have been interpreted in terms of magnetic polaron formation and coalescence. These features make EuB$_6$ fundamentally distinct from conventional localized-spin ferromagnets: its magnetic response near the transition is governed not simply by independent atomic moments, but by nanoscale, field-tunable correlated clusters. Here, we focus on the thermodynamic consequences of the magnetic polaron physics in EuB$_6$, and in particular on its impact on the magnetocaloric response. 

In this work, we show that EuB$_6$ exhibits an exceptional cryogenic magnetocaloric response that coincides with magnetic polaron formation. We find that EuB$_6$ displays giant isothermal entropy change $\Delta S_m$ (-42 J$\cdot$ kg$^{-1}\cdot$K$^{-1}$ under 5 T) and adiabatic temperature change $\Delta T_{ad}$ (19 K under 5 T) near its magnetic transition at 12 K. Benchmarking against representative magnetocaloric materials with Curie temperatures in the 10-40 K window reveals that EuB$_6$ simultaneously achieves among the largest entropy change and the largest adiabatic temperature change reported to date (Fig.~\ref{fig:fig1}b), while maintaining the intrinsic reversibility of a second-order transition. Through combined thermodynamic and magnetic analysis, we show that this remarkable performance correlates with an intermediate magnetic length scale associated with nanoscale ferromagnetic droplets. These results identify magnetic polarons as a previously unrecognized route to high-performance cryogenic magnetocaloric cooling and establish an emergent design principle for discovering magnetocaloric materials beyond conventional localized-spin and first-order-transition paradigms.

\section{Results and Discussions}
\subsection{Thermodynamic characterization and magnetocaloric response}

Details of the experimental methods are provided in the Supplementary Information. Briefly, EuB$_6$ powders were synthesized by a solid-state reaction using europium oxide (Eu$_2$O$_3$) and boron. Unlike most previous studies that grew EuB$_6$ single crystals using an Al flux~\cite{Sullow.1998.Structure}, we used a high-pressure laser floating zone technique~\cite{gomez2024advances} to grow an EuB$_6$ single crystal from the prereacted powders. EuB$_6$ has a cubic CaB$_6$-type crystal structure with space group \textit{Pm3m} (space group number 221)~\cite{Sullow.1998.Structure}. A schematic of its crystal structure is shown in Fig.~S1, where Eu atoms sit at the corners of a cubic lattice and six B atoms form an octahedral cage inside the unit cell. A small magnetic anisotropy with low magnetic fields below 0.5 T was noted previously~\cite{Sullow.1998.Structure}, which is not expected to significantly affect the magnetocaloric properties under strong fields. 

\begin{figure}[!htb]
\includegraphics[width=\textwidth]{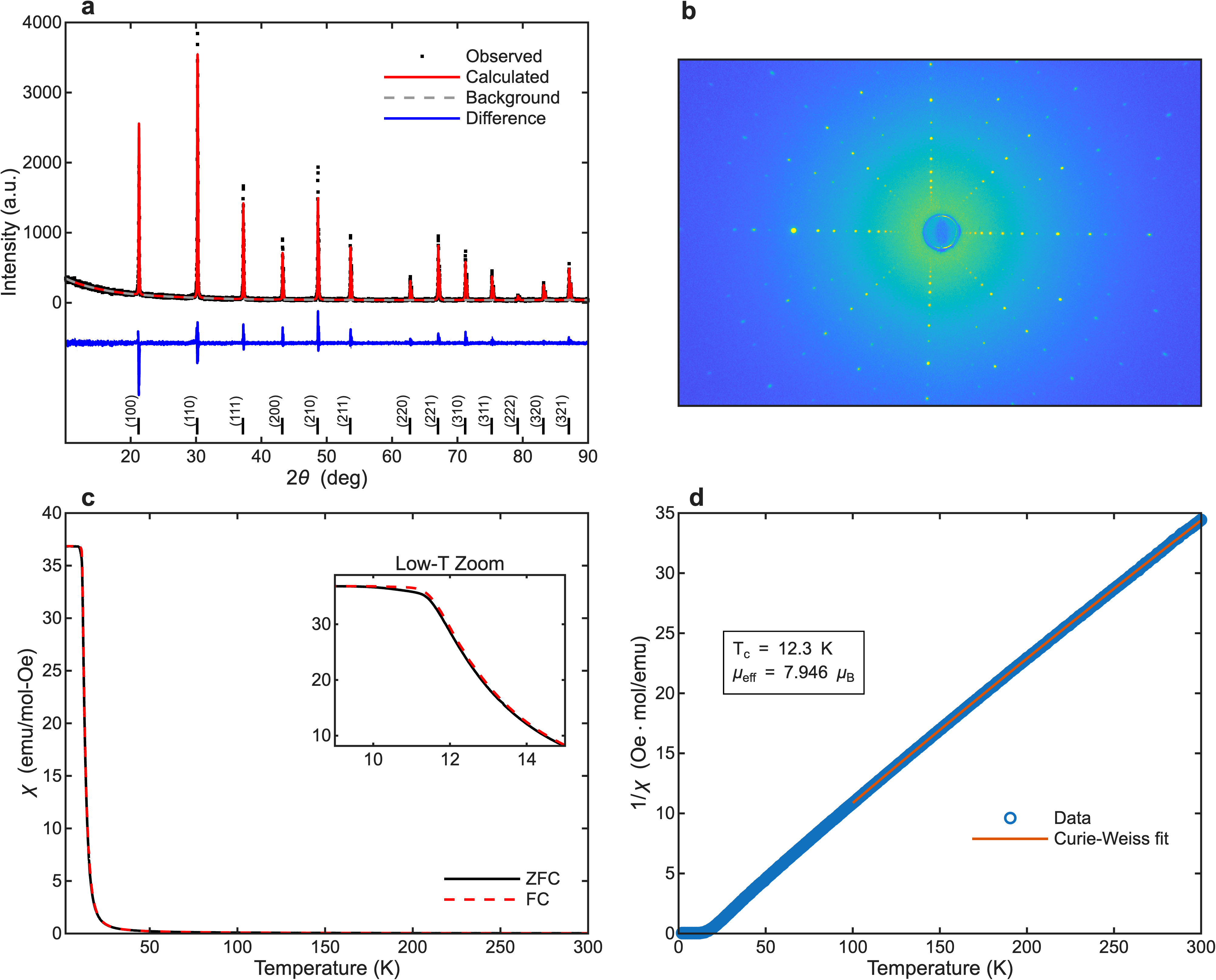}
\caption{\textbf{Structural and magnetic characterization of EuB$_6$.} \textbf{a}, Powder X-ray diffraction pattern with Rietveld refinement. Black symbols represent the observed data, the red line indicates the calculated profile, the gray dashed line shows the background, and the blue line is the difference curve. Vertical tick marks denote the Bragg peak positions. \textbf{b}, Back-reflection Laue diffraction pattern of a EuB$_6$ single crystal. \textbf{c}, Temperature dependence of the magnetic susceptibility measured under an applied field of 100 Oe for zero-field-cooled (ZFC) and field-cooled (FC) conditions. The inset shows a zoomed view near the ferromagnetic transition. \textbf{d}, Inverse magnetic susceptibility as a function of temperature. The solid line represents a Curie–Weiss fit, yielding a Curie temperature $T_C$ = 12.3 K and an effective magnetic moment of 7.946 $\mu_B$.} 
\label{fig:fig2}
\end{figure}

To establish the structural quality and intrinsic magnetic behavior of our EuB$_6$ crystal, we performed comprehensive structural and magnetic characterizations, as summarized in Fig.~\ref{fig:fig2}. The crystal structure and phase purity of EuB$_6$ were examined by powder X-ray diffraction (XRD) with Rietveld refinement (Fig.~\ref{fig:fig2}a). The diffraction pattern is well described by a cubic structure with space group \textit{Pm3m}, yielding a lattice parameter of 4.1851 \AA, consistent with previous reports~\cite{kasaya1978study}. 
The high crystalline quality of the sample is further confirmed by the back-reflection Laue diffraction pattern (Fig.~\ref{fig:fig2}b), which exhibits sharp diffraction spots with clear fourfold symmetry. The low-field magnetic susceptibility (Fig.~\ref{fig:fig2}c) exhibits a sharp increase near 12 K, consistent with the onset of ferromagnetic ordering~\cite{gao2021time}. A slight irreversibility between zero-field-cooled (ZFC) and field-cooled (FC) curves is observed below the transition (shown in the inset), which can be attributed to weak domain effects. The inverse susceptibility is plotted as a function of temperature in Fig.~\ref{fig:fig2}d. At high temperatures, the data follow a linear Curie-Weiss behavior. A linear fit yields a Curie temperature of 12.3 K and an effective magnetic moment of 7.946 $\mu_B$ per Eu$^{2+}$ ($\mu_B$ is the Bohr magneton), which is consistent with previous reports and close to the expected value for the local $4f^7$ orbital (7.94 $\mu_B$)~\cite{gao2021time}. The extracted $T_c$ is consistent with previous reports~\cite{degiorgi1997low,Sullow.1998.Structure}.

To quantify the magnetocaloric response, we conducted systematic thermodynamic and magnetization measurements. Figure S5 in the Supplementary Information shows the temperature-dependent specific heat measured under magnetic fields from 0 to 5 T, together with previously reported data below 20 K~\cite{Sullow.1998.Structure}. At zero field, a pronounced peak is observed near $T_c \approx$ 12 K, consistent with the ferromagnetic transition. With increasing magnetic field, the peak progressively broadens and shifts to higher temperature. Below 20 K, our measurements are in good overall agreement with previously reported data~\cite{Sullow.1998.Structure}. The total entropy $S(T,H)$, obtained by integrating the specific heat $C_p$ over temperature
\begin{equation}
S(T,H)=\int_0^T\frac{C_p(T',H)}{T'}dT',
\end{equation}
is shown in Fig.~\ref{fig:fig3}a (see Supplementary Note 2 for the extrapolation of specific heat to below 2 K). The entropy curves display a clear field dependence in the vicinity of the magnetic transition, while converging at higher temperatures as expected when magnetic contributions become negligible. From $S(T,H)$, $\Delta S_M$ and $\Delta T_{ad}$ can be determined as illustrated by the arrows in Fig.~\ref{fig:fig3}a. The isothermal entropy change $\Delta S_M$ can be derived from both specific heat and magnetization measurements. Using thermodynamic Maxwell relations, $\Delta S_M$ can be determined from the magnetization data~\cite{pecharsky1999magnetocaloric}:
\begin{equation} \label{eq:entropy}
\Delta S_M = \mu_B \int_{H_i}^{H_f} 
\left( \frac{\partial M}{\partial T} \right)_{H} \, dH,
\end{equation}
where $M$ is the magnetization, and $H_i$ and $H_f$ are the initial and final magnetic fields, respectively. $\Delta S_M$ determined from specific heat is shown as solid lines in Fig.~\ref{fig:fig3}b, while $\Delta S_M$ determined from magnetization is shown as dashed lines. The two independent methods show good overall agreement, especially under strong fields. Under 5 T, the peak $-\Delta S_M$ reaches around 42\,$\mathrm{J\,kg^{-1}K^{-1}}$ near the transition temperature, which is the highest reported experimental value in this temperature range. Notably, the entropy change remains substantial over a relatively wide temperature window, which is advantageous for practical cooling applications. 

Figure~\ref{fig:fig3}c shows the adiabatic temperature change $\Delta T_{ad}$ calculated from the specific heat data. The peak $\Delta T_{ad}$ value reaches 19 K under 5\,T with an initial temperature around 12 K. This value is significantly higher than other reported materials in this temperature range and also much higher than our previous atomistic spin dynamics (ASD) prediction of 8.1 K~\cite{Yuan2024enhancing}, suggesting that the ASD simulation missed a certain key mechanism in the paramagnetic phase from 12 K to around 30 K. As will be discussed in detail later, this range coincides with the temperature window in which the magnetic polarons are present and active in EuB$_6$~~\cite{snow2001magnetic,brooks2004magnetic,zhang2009nonlinear,beaudin2026observation}. The same measurement and analysis were conducted for EuB$_6$ powder (see Supplementary Note 3), with peak $-\Delta S_M$ and $\Delta T_{ad}$ under 5 T reaching 38\,$\mathrm{J\,kg^{-1}K^{-1}}$ and 16 K, respectively. To further verify the large $\Delta T_{ad}$, we performed a direct temperature measurement of a large polycrystalline sample under a magnetic field ramping up at a rate of 200 Oe/s. The sample was initially maintained at 12 K and well thermally insulated with the only heat loss through the thermocouple wires (more details in Supplementary Methods). We measured a direct temperature rise of 11 K when the field was ramped up to 5 T, as shown in Fig.~\ref{fig:fig3}d. This value is smaller than $\Delta T_{ad}$ due to the finite field ramping rate and the heat loss during the ramping process. Using a thermal model to fit the temperature evolution (see Supplementary Note 4), we extracted a $\Delta T_{ad}$ value of 15.1 K, which is close to that of the powder sample determined thermodynamically. This result suggests that EuB$_6$ exhibits one of the largest adiabatic temperature changes reported under a magnetic field of 5 T, rivaling or exceeding many classical giant magnetocaloric materials, while maintaining the intrinsic reversibility of a second-order transition. 

Taken together, these results establish that EuB$_6$ exhibits an unusually strong magnetocaloric response in the 10-30 K regime. In particular, it simultaneously achieves a large isothermal entropy change and a large adiabatic temperature change under moderate magnetic fields, a combination that is rarely realized in materials with second-order magnetic transitions. Moreover, the entropy change remains substantial over a relatively broad temperature window above $T_c$, and the adiabatic temperature change significantly exceeds expectations based on conventional paramagnetic scaling and ASD simulation with a Heisenberg Hamiltonian with fixed local spin moments~\cite{Yuan2024enhancing}. These features indicate that the magnetocaloric response in EuB$_6$ cannot be fully understood within a simple localized-spin picture, and instead point to an additional mechanism for the unusual enhancement.

\begin{figure}[!htb]
\includegraphics[width=\textwidth]{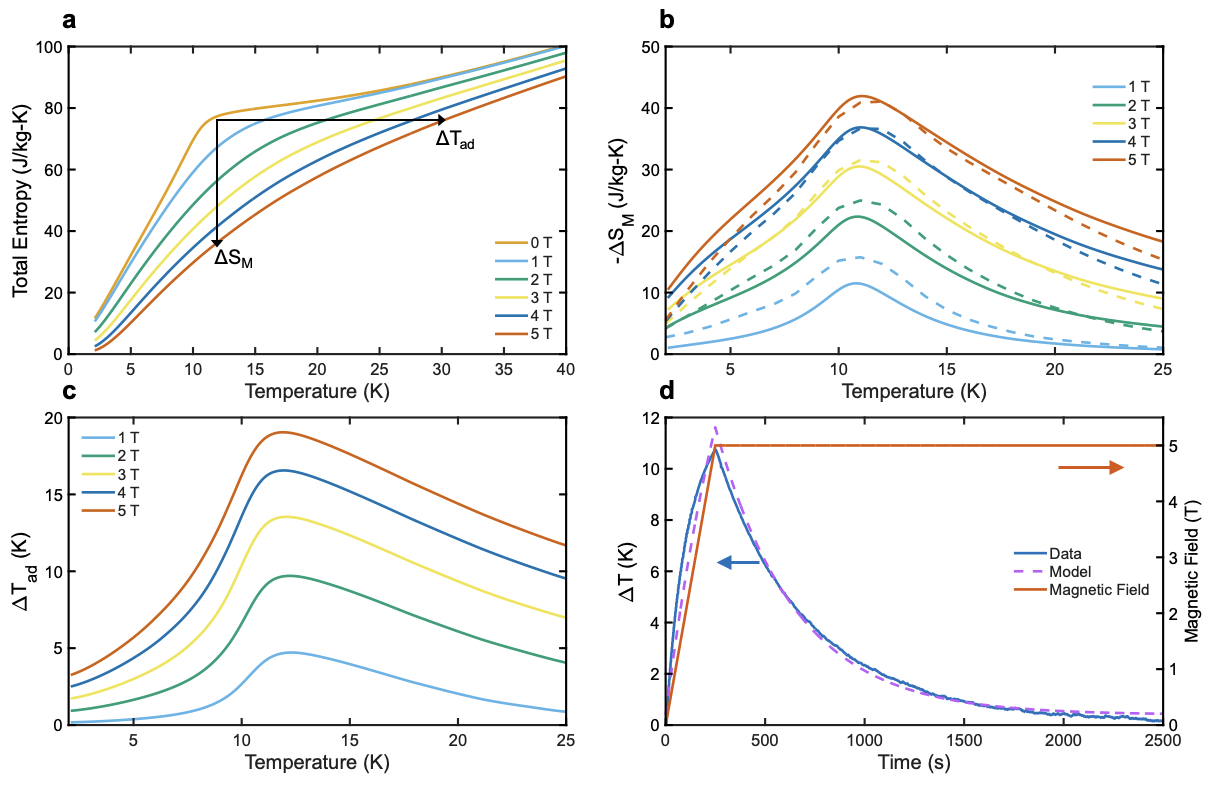}
\caption{\textbf{Thermodynamic and direct measurements of the magnetocaloric response in EuB$_6$.} \textbf{a}, Total entropy $S(T,H)$ derived from specific heat measurements under magnetic fields from 0 to 5 T. The separation between entropy curves at different fields defines the isothermal entropy change $\Delta S_M$, while the horizontal displacement at constant entropy corresponds to the adiabatic temperature change $\Delta T_{ad}$. \textbf{b}, Isothermal entropy change 
$-\Delta S_M$ as a function of temperature for fields up to 5 T. Solid lines are obtained from specific heat, while dashed lines are derived from magnetization via the Maxwell relation. \textbf{c}, Adiabatic temperature change $\Delta T_{ad}$ calculated from the specific heat data. \textbf{d}, Direct measurement of the transient temperature response of a large polycrystalline sample under an applied magnetic field. The experimental data (blue) are compared with a model fit (purple dashed), while the magnetic field profile (orange, right axis) is shown simultaneously.} 
\label{fig:fig3}
\end{figure}

\subsection{Magnetic polarons as the origin of the enhanced magnetocaloric response}

While the large magnetocaloric effect in EuB$_6$ can be naturally anticipated from the $4f^7$ configuration of Eu$^{2+}$ ions, which provides a large spin entropy (spin moment $S$ = 7/2) with minimal crystal-field effects (orbital moment $L$ = 0), a detailed analysis reveals that this single-ion picture is insufficient to explain the observed magnitude and field dependence. Instead, a consistent set of thermodynamic, magnetic, and transport signatures points to the formation of magnetic polarons as the key microscopic origin.

\begin{figure}[!htb]
\includegraphics[width=\textwidth]{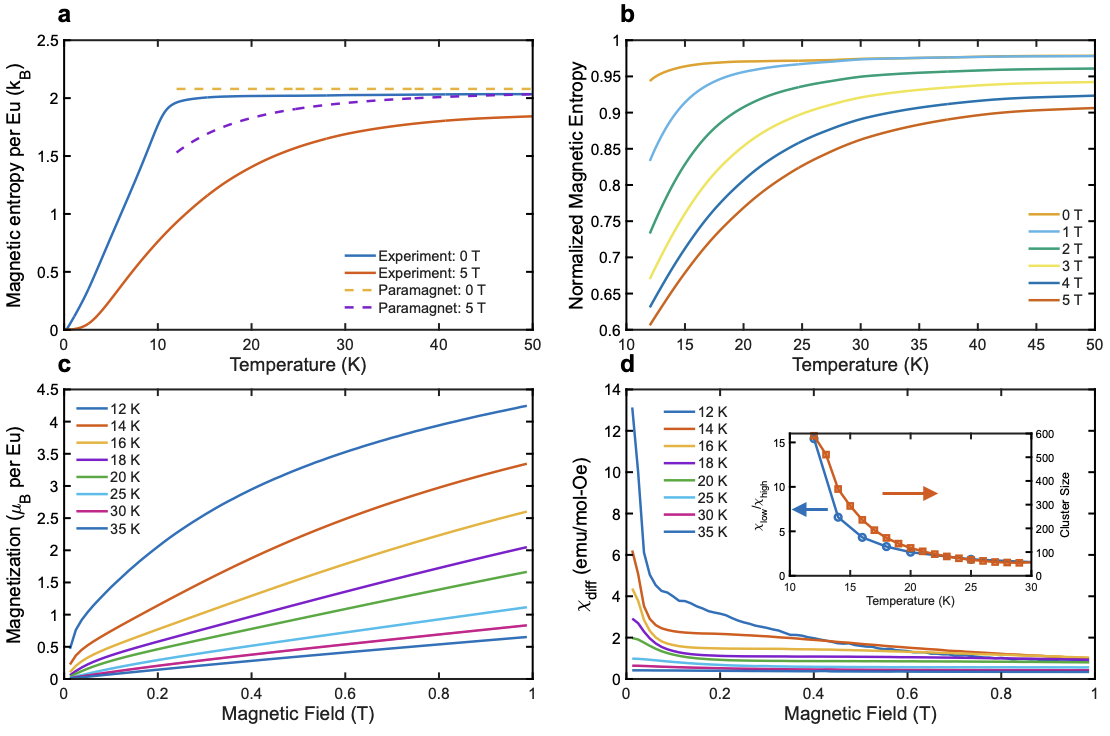}
\caption{\textbf{Signatures of magnetic polaron formation in EuB$_6$.} \textbf{a,} Magnetic entropy per Eu extracted from specific heat after subtracting phonon and electronic contributions, in comparison to a simple paramagnet.
\textbf{b,} Magnetic entropy normalized to that of an ideal paramagnet as a function of temperature and magnetic field. 
\textbf{c,} Low-field magnetization at selected temperatures, showing strongly nonlinear field dependence in the vicinity of $T_c$, in contrast to the linear response expected for a simple paramagnet.
\textbf{d,} Differential susceptibility $\chi_{\mathrm{diff}}$ as a function of magnetic field, revealing a strong low-field enhancement especially near $T_c$. \textbf{Inset:} Temperature dependence of the ratio between low-field and high-field susceptibility (left axis) and the effective cluster size (number of Eu$^{2+}$ ions) extracted from a Langevin-type model (right axis).} 
\label{fig:fig4}
\end{figure}

The magnetic entropy, extracted from specific heat after subtracting phonon and electronic contributions (see Supplementary Note 1), already shows clear deviations from simple paramagnetic behavior. As shown in Fig.~\ref{fig:fig4}a, the zero-field entropy approaches the paramagnetic limit at high temperature, while the entropy under an applied field of 5 T is strongly suppressed below the paramagnetic model (see Supplementary Note 5 for details of the paramagnetic reference model), far exceeding expectations for non-interacting spins. This anomalous field sensitivity is further evident in the normalized magnetic entropy (Fig.~\ref{fig:fig4}b), where the experimental magnetic entropy is normalized by the paramagnetic value at the same temperature and magnetic field. Figure ~\ref{fig:fig4}b exhibits a pronounced reduction near and above the Curie temperature $T_c$, indicating that the magnetic degrees of freedom are not independent but instead form collective entities that respond strongly to external fields.

Consistent with this picture, the low-field magnetization exhibits pronounced nonlinearity from 12 to 20 K (Fig.~\ref{fig:fig4}c), while the differential susceptibility shows a strong low-field enhancement that rapidly decreases with increasing field (Fig.~\ref{fig:fig4}d). The ratio of low-field (0.1 to 0.5 T) to high-field (1.5 to 2.5 T) susceptibility peaks above 15 near $T_c$. This enhanced low-field response is consistent with the established magnetic polaron picture, where nanometer-scale spin clusters are formed through carrier-mediated interactions. To provide a rough phenomenological parameterization of the polaron picture, we used a simple Langevin-type model that encodes the magnetic response of spin clusters~\cite{bansmann2005magnetic} to fit the low-field magnetization data and extract the effective spin cluster size (see Supplementary Note 6). The extracted effective cluster size is shown in the inset of Fig.~\ref{fig:fig4}d and, in this approximation, the cluster size necessarily diverges as $T_c$ is approached. Close to $T_c$, the extracted effective cluster size reaches $\sim 600$ Eu spins, corresponding to a characteristic length scale of $\sim 4.5$ nm, consistent with magnetic correlations freezing within the intermediate regime between atomic spins and mesoscopic magnetic particles (Fig.~\ref{fig:fig1}a). Notably, the characteristic length scale inferred from our Langevin analysis is smaller than the spin-correlation length reported by neutron scattering ($\sim$10 nm near 13 K)~\cite{beaudin2026observation}. This is likely due to the overly simple parameterization the Langevin fit yields within a noninteracting-cluster picture, and our magnetization data's inability to isolate these clusters from any individual moments polarized via a ferromagnetic mean field close to $T_c$. As an illustration of an uncorrelated extreme without clusters, a simple Brillouin-type description of independent Eu$^{2+}$ moments fails to reproduce the pronounced low-field curvature of the magnetization curves (see Supplementary Note 6), particularly near $T_c$, suggesting that the relevant field-responsive objects are not isolated spins but more likely correlated magnetic clusters with much larger effective moments that are consistent with prior magnetic polaron interpretations. These results point to highly field-sensitive nanoscale magnetic clusters that are responsible for the enhanced magnetocaloric response.

Previous studies have established multiple signatures of magnetic polaron formation in EuB$_6$ above $T_c$, including Raman scattering~\cite{snow2001magnetic}, muon spin rotation~\cite{brooks2004magnetic}, magnetotransport~\cite{sullow2000metallization,guy1980charge,zhang2009nonlinear}, and neutron scattering~\cite{beaudin2026observation}. As a consistency check, we also conducted magnetotransport measurement of our EuB$_6$ sample and observed a large negative magnetoresistance above $T_c$ that is consistent with previous reports (see Supplementary Note 7). The coincidence of large negative magnetoresistance, strongly enhanced differential susceptibility, and anomalously suppressed magnetic entropy by magnetic fields above $T_c$ indicates that magnetic polaron formation provides a consistent framework linking both transport and magnetocaloric responses in EuB$_6$. In this picture, charge carriers locally polarize surrounding spins to form nanoscale ferromagnetic regions above $T_c$, creating a heterogeneous magnetic state composed of fluctuating clusters embedded in a paramagnetic background~\cite{snow2001magnetic}. These clusters act as large effective moments that can be readily aligned by an external magnetic field while retaining a substantial configurational entropy because they are small in size and dynamically fluctuating. While critical magnetic fluctuations near $T_c$ may contribute to enhanced field responsiveness~\cite{hohenberg1977theory}, their influence is expected to be confined to a relatively narrow temperature window in an unfrustrated three-dimensional ferromagnet~\cite{beaudin2026observation}. In contrast, the combined thermodynamic, magnetic, and transport signatures observed here persist over a much broader temperature range above $T_c$, and are most consistently explained by the formation of magnetic polarons. This extended temperature window of enhanced field responsiveness provides a natural explanation for the large magnetocaloric response observed in this system.

Magnetic polaron formation is not unique to EuB$_6$, and has been extensively documented in low-carrier-density Eu-based magnetic semiconductors such as EuO and EuS~\cite{snow2001magnetic,storchak2010magnetic}. Similar cluster physics has also been invoked in manganites, where magnetic polarons and nanoscale phase separation are widely believed to underlie colossal magnetoresistance and strong field-dependent transport responses~\cite{emin2012magnetic,moreo2000giant,calderon2000stability}. However, in these systems the impact of magnetic polarons on thermodynamic quantities, particularly the magnetocaloric response, has remained largely unexplored or obscured by structural and first-order effects. The present results establish a direct link between magnetic polaron formation and enhanced magnetocaloric performance, suggesting that correlated magnetic clusters provide a general route for optimizing magnetocaloric materials. More broadly, this perspective establishes an intermediate magnetic length scale as a key design principle for next-generation magnetocaloric materials.

\section{Conclusion}

In summary, we demonstrate that EuB$_6$ exhibits an exceptional magnetocaloric response in the technologically important 10-40~K temperature range, characterized by simultaneously large isothermal entropy change and adiabatic temperature change under moderate magnetic fields. Through combined thermodynamic, magnetic, and transport measurements, we establish that this behavior is most consistently understood within the magnetic polaron regime, where nanoscale ferromagnetic clusters emerge above the Curie temperature and evolve into a percolating magnetic state upon cooling. These correlated entities provide an optimal balance between entropy density and field responsiveness, enabling an efficient conversion of magnetic entropy into temperature change. Our results identify magnetic polarons as a previously unrecognized mechanism for enhancing magnetocaloric performance and establish an intermediate magnetic length scale as a key design principle. More generally, this framework suggests that targeting systems with emergent, field-tunable magnetic clusters may provide a broadly applicable strategy for designing high-performance magnetocaloric materials. We note that similar nanoscale polar domains have been associated with enhanced electrocaloric responses in relaxor ferroelectric polymers~\cite{qian2021high}, hinting at a possible broader relevance of intermediate correlation length scales.

\begin{acknowledgments}
Measurements were done at UCSB Materials Research Laboratory Shared Experimental Facilities, which are supported by the NSF MRSEC Program under award number DMR-2308708.
\end{acknowledgments}

\section*{Data Availability}
All data will be available from a public online data repository upon publication of this work.

\section*{Funding}
This work is based on research partially supported by the National Aeronautics and Space Administration under award number 80NSSC21K1812 (for the development of cryogenic magnetocaloric materials) and a seed grant provided by Institute of Energy Efficiency at the University of California, Santa Barbara (UCSB) (for discovering new magnetic materials). C.B. also acknowledges the support of the National Science Foundation (NSF) Gradate Research Fellowship Program under the award number DMR-2139319.

\section*{Author Contributions}
B.L. and S.D.W initiated and supervised the research. J.A. and C.B. synthesized the sample and conducted structural, magnetization, and thermodynamic measurements. F.Z. conducted direct temperature measurement and magnetotransport measurement. F.Z., S.D.W and B.L. drafted the manuscript. All authors reviewed and edited the manuscript. 

\section*{Competing Interests}
A provisional United States patent application on the discovery in this paper has been filed.

\bibliography{references.bib}%

\begin{thebibliography}{10}%
\makeatletter
\providecommand \@ifxundefined [1]{%
 \@ifx{#1\undefined}
}%
\providecommand \@ifnum [1]{%
 \ifnum #1\expandafter \@firstoftwo
 \else \expandafter \@secondoftwo
 \fi
}%
\providecommand \@ifx [1]{%
 \ifx #1\expandafter \@firstoftwo
 \else \expandafter \@secondoftwo
 \fi
}%
\providecommand \natexlab [1]{#1}%
\providecommand \enquote  [1]{``#1''}%
\providecommand \bibnamefont  [1]{#1}%
\providecommand \bibfnamefont [1]{#1}%
\providecommand \citenamefont [1]{#1}%
\providecommand \href@noop [0]{\@secondoftwo}%
\providecommand \href [0]{\begingroup \@sanitize@url \@href}%
\providecommand \@href[1]{\@@startlink{#1}\@@href}%
\providecommand \@@href[1]{\endgroup#1\@@endlink}%
\providecommand \@sanitize@url [0]{\catcode `\\12\catcode `\$12\catcode
  `\&12\catcode `\#12\catcode `\^12\catcode `\_12\catcode `\%12\relax}%
\providecommand \@@startlink[1]{}%
\providecommand \@@endlink[0]{}%
\providecommand \url  [0]{\begingroup\@sanitize@url \@url }%
\providecommand \@url [1]{\endgroup\@href {#1}{\urlprefix }}%
\providecommand \urlprefix  [0]{URL }%
\providecommand \Eprint [0]{\href }%
\providecommand \doibase [0]{https://doi.org/}%
\providecommand \selectlanguage [0]{\@gobble}%
\providecommand \bibinfo  [0]{\@secondoftwo}%
\providecommand \bibfield  [0]{\@secondoftwo}%
\providecommand \translation [1]{[#1]}%
\providecommand \BibitemOpen [0]{}%
\providecommand \bibitemStop [0]{}%
\providecommand \bibitemNoStop [0]{.\EOS\space}%
\providecommand \EOS [0]{\spacefactor3000\relax}%
\providecommand \BibitemShut  [1]{\csname bibitem#1\endcsname}%
\let\auto@bib@innerbib\@empty
\bibitem [{\citenamefont {Bocarsly}\ \emph {et~al.}(2018)\citenamefont
  {Bocarsly}, \citenamefont {Need}, \citenamefont {Seshadri},\ and\
  \citenamefont {Wilson}}]{bocarsly2018magnetoentropic}%
  \BibitemOpen
  \bibfield  {author} {\bibinfo {author} {\bibfnamefont {J.~D.}\ \bibnamefont
  {Bocarsly}}, \bibinfo {author} {\bibfnamefont {R.~F.}\ \bibnamefont {Need}},
  \bibinfo {author} {\bibfnamefont {R.}~\bibnamefont {Seshadri}},\ and\
  \bibinfo {author} {\bibfnamefont {S.~D.}\ \bibnamefont {Wilson}},\ }\bibfield
   {title} {\bibinfo {title} {Magnetoentropic signatures of skyrmionic phase
  behavior in {FeGe}},\ }\href@noop {} {\bibfield  {journal} {\bibinfo
  {journal} {Physical Review B}\ }\textbf {\bibinfo {volume} {97}},\ \bibinfo
  {pages} {100404} (\bibinfo {year} {2018})}\BibitemShut {NoStop}%
\bibitem [{\citenamefont {Burns}\ \emph {et~al.}(1993)\citenamefont {Burns},
  \citenamefont {Scroger}, \citenamefont {Strouse}, \citenamefont {Croarkin},\
  and\ \citenamefont {Guthrie}}]{burns1993temperature}%
  \BibitemOpen
  \bibfield  {author} {\bibinfo {author} {\bibfnamefont {G.~W.}\ \bibnamefont
  {Burns}}, \bibinfo {author} {\bibfnamefont {M.~G.}\ \bibnamefont {Scroger}},
  \bibinfo {author} {\bibfnamefont {G.~F.}\ \bibnamefont {Strouse}}, \bibinfo
  {author} {\bibfnamefont {M.~C.}\ \bibnamefont {Croarkin}},\ and\ \bibinfo
  {author} {\bibfnamefont {W.~F.}\ \bibnamefont {Guthrie}},\ }\href@noop {}
  {\emph {\bibinfo {title} {Temperature-Electromotive Force Reference Functions
  and Tables for the Letter-Designated Thermocouple Types Based on the
  ITS-90}}},\ \bibinfo {type} {Tech. Rep.}\ \bibinfo {number} {NIST Monograph
  175}\ (\bibinfo  {institution} {National Institute of Standards and
  Technology},\ \bibinfo {address} {Gaithersburg, MD},\ \bibinfo {year}
  {1993})\BibitemShut {NoStop}%
\bibitem [{\citenamefont {Sample}\ \emph {et~al.}(1974)\citenamefont {Sample},
  \citenamefont {Neuringer},\ and\ \citenamefont {Rubin}}]{sample1974low}%
  \BibitemOpen
  \bibfield  {author} {\bibinfo {author} {\bibfnamefont {H.}~\bibnamefont
  {Sample}}, \bibinfo {author} {\bibfnamefont {L.}~\bibnamefont {Neuringer}},\
  and\ \bibinfo {author} {\bibfnamefont {L.}~\bibnamefont {Rubin}},\ }\bibfield
   {title} {\bibinfo {title} {Low temperature thermometry in high magnetic
  fields. {III. Carbon resistors (0.5--4.2 K); thermocouples.}},\ }\href@noop
  {} {\bibfield  {journal} {\bibinfo  {journal} {Review of Scientific
  Instruments}\ }\textbf {\bibinfo {volume} {45}},\ \bibinfo {pages} {64}
  (\bibinfo {year} {1974})}\BibitemShut {NoStop}%
\bibitem [{\citenamefont {Süllow}\ \emph {et~al.}(1998)\citenamefont
  {Süllow}, \citenamefont {Prasad}, \citenamefont {Aronson}, \citenamefont
  {Sarrao}, \citenamefont {Fisk}, \citenamefont {Hristova}, \citenamefont
  {Lacerda}, \citenamefont {Hundley}, \citenamefont {Vigliante},\ and\
  \citenamefont {Gibbs}}]{Sullow.1998.Structure}%
  \BibitemOpen
  \bibfield  {author} {\bibinfo {author} {\bibfnamefont {S.}~\bibnamefont
  {Süllow}}, \bibinfo {author} {\bibfnamefont {I.}~\bibnamefont {Prasad}},
  \bibinfo {author} {\bibfnamefont {M.~C.}\ \bibnamefont {Aronson}}, \bibinfo
  {author} {\bibfnamefont {J.~L.}\ \bibnamefont {Sarrao}}, \bibinfo {author}
  {\bibfnamefont {Z.}~\bibnamefont {Fisk}}, \bibinfo {author} {\bibfnamefont
  {D.}~\bibnamefont {Hristova}}, \bibinfo {author} {\bibfnamefont {A.~H.}\
  \bibnamefont {Lacerda}}, \bibinfo {author} {\bibfnamefont {M.~F.}\
  \bibnamefont {Hundley}}, \bibinfo {author} {\bibfnamefont {A.}~\bibnamefont
  {Vigliante}},\ and\ \bibinfo {author} {\bibfnamefont {D.}~\bibnamefont
  {Gibbs}},\ }\bibfield  {title} {\bibinfo {title} {{Structure and magnetic
  order of {EuB$_6$}}},\ }\href {https://doi.org/10.1103/physrevb.57.5860}
  {\bibfield  {journal} {\bibinfo  {journal} {Physical Review B}\ }\textbf
  {\bibinfo {volume} {57}},\ \bibinfo {pages} {5860} (\bibinfo {year}
  {1998})}\BibitemShut {NoStop}%
\bibitem [{\citenamefont {{Franco}}\ \emph {et~al.}(2012)\citenamefont
  {{Franco}}, \citenamefont {{Bl{\'a}zquez}}, \citenamefont {{Ingale}},\ and\
  \citenamefont {{Conde}}}]{Franco2012magnetocaloric}%
  \BibitemOpen
  \bibfield  {author} {\bibinfo {author} {\bibfnamefont {V.}~\bibnamefont
  {{Franco}}}, \bibinfo {author} {\bibfnamefont {J.~S.}\ \bibnamefont
  {{Bl{\'a}zquez}}}, \bibinfo {author} {\bibfnamefont {B.}~\bibnamefont
  {{Ingale}}},\ and\ \bibinfo {author} {\bibfnamefont {A.}~\bibnamefont
  {{Conde}}},\ }\bibfield  {title} {\bibinfo {title} {{The magnetocaloric
  effect and magnetic refrigeration near room temperature: materials and
  models}},\ }\href {https://doi.org/10.1146/annurev-matsci-062910-100356}
  {\bibfield  {journal} {\bibinfo  {journal} {Annual Review of Materials
  Research}\ }\textbf {\bibinfo {volume} {42}},\ \bibinfo {pages} {305}
  (\bibinfo {year} {2012})}\BibitemShut {NoStop}%
\bibitem [{\citenamefont {Bansmann}\ \emph {et~al.}(2005)\citenamefont
  {Bansmann}, \citenamefont {Baker}, \citenamefont {Binns}, \citenamefont
  {Blackman}, \citenamefont {Bucher}, \citenamefont {Dorantes-D{\'a}vila},
  \citenamefont {Dupuis}, \citenamefont {Favre}, \citenamefont {Kechrakos},
  \citenamefont {Kleibert} \emph {et~al.}}]{bansmann2005magnetic}%
  \BibitemOpen
  \bibfield  {author} {\bibinfo {author} {\bibfnamefont {J.}~\bibnamefont
  {Bansmann}}, \bibinfo {author} {\bibfnamefont {S.}~\bibnamefont {Baker}},
  \bibinfo {author} {\bibfnamefont {C.}~\bibnamefont {Binns}}, \bibinfo
  {author} {\bibfnamefont {J.}~\bibnamefont {Blackman}}, \bibinfo {author}
  {\bibfnamefont {J.-P.}\ \bibnamefont {Bucher}}, \bibinfo {author}
  {\bibfnamefont {J.}~\bibnamefont {Dorantes-D{\'a}vila}}, \bibinfo {author}
  {\bibfnamefont {V.}~\bibnamefont {Dupuis}}, \bibinfo {author} {\bibfnamefont
  {L.}~\bibnamefont {Favre}}, \bibinfo {author} {\bibfnamefont
  {D.}~\bibnamefont {Kechrakos}}, \bibinfo {author} {\bibfnamefont
  {A.}~\bibnamefont {Kleibert}}, \emph {et~al.},\ }\bibfield  {title} {\bibinfo
  {title} {Magnetic and structural properties of isolated and assembled
  clusters},\ }\href@noop {} {\bibfield  {journal} {\bibinfo  {journal}
  {Surface Science Reports}\ }\textbf {\bibinfo {volume} {56}},\ \bibinfo
  {pages} {189} (\bibinfo {year} {2005})}\BibitemShut {NoStop}%
\bibitem [{\citenamefont {Sakamoto}\ \emph {et~al.}(2016)\citenamefont
  {Sakamoto}, \citenamefont {Anh}, \citenamefont {Hai}, \citenamefont
  {Shibata}, \citenamefont {Takeda}, \citenamefont {Kobayashi}, \citenamefont
  {Takahashi}, \citenamefont {Koide}, \citenamefont {Tanaka},\ and\
  \citenamefont {Fujimori}}]{sakamoto2016magnetization}%
  \BibitemOpen
  \bibfield  {author} {\bibinfo {author} {\bibfnamefont {S.}~\bibnamefont
  {Sakamoto}}, \bibinfo {author} {\bibfnamefont {L.~D.}\ \bibnamefont {Anh}},
  \bibinfo {author} {\bibfnamefont {P.~N.}\ \bibnamefont {Hai}}, \bibinfo
  {author} {\bibfnamefont {G.}~\bibnamefont {Shibata}}, \bibinfo {author}
  {\bibfnamefont {Y.}~\bibnamefont {Takeda}}, \bibinfo {author} {\bibfnamefont
  {M.}~\bibnamefont {Kobayashi}}, \bibinfo {author} {\bibfnamefont
  {Y.}~\bibnamefont {Takahashi}}, \bibinfo {author} {\bibfnamefont
  {T.}~\bibnamefont {Koide}}, \bibinfo {author} {\bibfnamefont
  {M.}~\bibnamefont {Tanaka}},\ and\ \bibinfo {author} {\bibfnamefont
  {A.}~\bibnamefont {Fujimori}},\ }\bibfield  {title} {\bibinfo {title}
  {Magnetization process of the n-type ferromagnetic semiconductor
  {(In,Fe)As:Be} studied by x-ray magnetic circular dichroism},\ }\href@noop {}
  {\bibfield  {journal} {\bibinfo  {journal} {Physical Review B}\ }\textbf
  {\bibinfo {volume} {93}},\ \bibinfo {pages} {035203} (\bibinfo {year}
  {2016})}\BibitemShut {NoStop}%
\bibitem [{\citenamefont {S{\"u}llow}\ \emph {et~al.}(2000)\citenamefont
  {S{\"u}llow}, \citenamefont {Prasad}, \citenamefont {Aronson}, \citenamefont
  {Bogdanovich}, \citenamefont {Sarrao},\ and\ \citenamefont
  {Fisk}}]{sullow2000metallization}%
  \BibitemOpen
  \bibfield  {author} {\bibinfo {author} {\bibfnamefont {S.}~\bibnamefont
  {S{\"u}llow}}, \bibinfo {author} {\bibfnamefont {I.}~\bibnamefont {Prasad}},
  \bibinfo {author} {\bibfnamefont {M.}~\bibnamefont {Aronson}}, \bibinfo
  {author} {\bibfnamefont {S.}~\bibnamefont {Bogdanovich}}, \bibinfo {author}
  {\bibfnamefont {J.}~\bibnamefont {Sarrao}},\ and\ \bibinfo {author}
  {\bibfnamefont {Z.}~\bibnamefont {Fisk}},\ }\bibfield  {title} {\bibinfo
  {title} {Metallization and magnetic order in {EuB$_6$}},\ }\href@noop {}
  {\bibfield  {journal} {\bibinfo  {journal} {Physical Review B}\ }\textbf
  {\bibinfo {volume} {62}},\ \bibinfo {pages} {11626} (\bibinfo {year}
  {2000})}\BibitemShut {NoStop}%
\bibitem [{\citenamefont {Guy}\ \emph {et~al.}(1980)\citenamefont {Guy},
  \citenamefont {Von~Molnar}, \citenamefont {Etourneau},\ and\ \citenamefont
  {Fisk}}]{guy1980charge}%
  \BibitemOpen
  \bibfield  {author} {\bibinfo {author} {\bibfnamefont {C.}~\bibnamefont
  {Guy}}, \bibinfo {author} {\bibfnamefont {S.}~\bibnamefont {Von~Molnar}},
  \bibinfo {author} {\bibfnamefont {J.}~\bibnamefont {Etourneau}},\ and\
  \bibinfo {author} {\bibfnamefont {Z.}~\bibnamefont {Fisk}},\ }\bibfield
  {title} {\bibinfo {title} {Charge transport and pressure dependence of
  {T$_c$} of single crystal, ferromagnetic {EuB$_6$}},\ }\href@noop {}
  {\bibfield  {journal} {\bibinfo  {journal} {Solid State Communications}\
  }\textbf {\bibinfo {volume} {33}},\ \bibinfo {pages} {1055} (\bibinfo {year}
  {1980})}\BibitemShut {NoStop}%
\bibitem [{\citenamefont {Zhang}\ \emph {et~al.}(2009)\citenamefont {Zhang},
  \citenamefont {Yu}, \citenamefont {von Moln{\'a}r}, \citenamefont {Fisk},\
  and\ \citenamefont {Xiong}}]{zhang2009nonlinear}%
  \BibitemOpen
  \bibfield  {author} {\bibinfo {author} {\bibfnamefont {X.}~\bibnamefont
  {Zhang}}, \bibinfo {author} {\bibfnamefont {L.}~\bibnamefont {Yu}}, \bibinfo
  {author} {\bibfnamefont {S.}~\bibnamefont {von Moln{\'a}r}}, \bibinfo
  {author} {\bibfnamefont {Z.}~\bibnamefont {Fisk}},\ and\ \bibinfo {author}
  {\bibfnamefont {P.}~\bibnamefont {Xiong}},\ }\bibfield  {title} {\bibinfo
  {title} {Nonlinear {H}all effect as a signature of electronic phase
  separation in the semimetallic ferromagnet {EuB$_6$}},\ }\href@noop {}
  {\bibfield  {journal} {\bibinfo  {journal} {Physical Review Letters}\
  }\textbf {\bibinfo {volume} {103}},\ \bibinfo {pages} {106602} (\bibinfo
  {year} {2009})}\BibitemShut {NoStop}%
\end{thebibliography}%


\begin{thebibliography}{44}%
\makeatletter
\providecommand \@ifxundefined [1]{%
 \@ifx{#1\undefined}
}%
\providecommand \@ifnum [1]{%
 \ifnum #1\expandafter \@firstoftwo
 \else \expandafter \@secondoftwo
 \fi
}%
\providecommand \@ifx [1]{%
 \ifx #1\expandafter \@firstoftwo
 \else \expandafter \@secondoftwo
 \fi
}%
\providecommand \natexlab [1]{#1}%
\providecommand \enquote  [1]{``#1''}%
\providecommand \bibnamefont  [1]{#1}%
\providecommand \bibfnamefont [1]{#1}%
\providecommand \citenamefont [1]{#1}%
\providecommand \href@noop [0]{\@secondoftwo}%
\providecommand \href [0]{\begingroup \@sanitize@url \@href}%
\providecommand \@href[1]{\@@startlink{#1}\@@href}%
\providecommand \@@href[1]{\endgroup#1\@@endlink}%
\providecommand \@sanitize@url [0]{\catcode `\\12\catcode `\$12\catcode `\&12\catcode `\#12\catcode `\^12\catcode `\_12\catcode `\%12\relax}%
\providecommand \@@startlink[1]{}%
\providecommand \@@endlink[0]{}%
\providecommand \url  [0]{\begingroup\@sanitize@url \@url }%
\providecommand \@url [1]{\endgroup\@href {#1}{\urlprefix }}%
\providecommand \urlprefix  [0]{URL }%
\providecommand \Eprint [0]{\href }%
\providecommand \doibase [0]{https://doi.org/}%
\providecommand \selectlanguage [0]{\@gobble}%
\providecommand \bibinfo  [0]{\@secondoftwo}%
\providecommand \bibfield  [0]{\@secondoftwo}%
\providecommand \translation [1]{[#1]}%
\providecommand \BibitemOpen [0]{}%
\providecommand \bibitemStop [0]{}%
\providecommand \bibitemNoStop [0]{.\EOS\space}%
\providecommand \EOS [0]{\spacefactor3000\relax}%
\providecommand \BibitemShut  [1]{\csname bibitem#1\endcsname}%
\let\auto@bib@innerbib\@empty
\bibitem [{\citenamefont {Al~Ghafri}\ \emph {et~al.}(2022)\citenamefont {Al~Ghafri}, \citenamefont {Munro}, \citenamefont {Cardella}, \citenamefont {Funke}, \citenamefont {Notardonato}, \citenamefont {Trusler}, \citenamefont {Leachman}, \citenamefont {Span}, \citenamefont {Kamiya}, \citenamefont {Pearce}, \citenamefont {Swanger}, \citenamefont {Rodriguez}, \citenamefont {Bajada}, \citenamefont {Jiao}, \citenamefont {Peng}, \citenamefont {Siahvashi}, \citenamefont {Johns},\ and\ \citenamefont {May}}]{Ghafri2022hydrogen}%
  \BibitemOpen
  \bibfield  {author} {\bibinfo {author} {\bibfnamefont {S.~Z.}\ \bibnamefont {Al~Ghafri}}, \bibinfo {author} {\bibfnamefont {S.}~\bibnamefont {Munro}}, \bibinfo {author} {\bibfnamefont {U.}~\bibnamefont {Cardella}}, \bibinfo {author} {\bibfnamefont {T.}~\bibnamefont {Funke}}, \bibinfo {author} {\bibfnamefont {W.}~\bibnamefont {Notardonato}}, \bibinfo {author} {\bibfnamefont {J.~P.~M.}\ \bibnamefont {Trusler}}, \bibinfo {author} {\bibfnamefont {J.}~\bibnamefont {Leachman}}, \bibinfo {author} {\bibfnamefont {R.}~\bibnamefont {Span}}, \bibinfo {author} {\bibfnamefont {S.}~\bibnamefont {Kamiya}}, \bibinfo {author} {\bibfnamefont {G.}~\bibnamefont {Pearce}}, \bibinfo {author} {\bibfnamefont {A.}~\bibnamefont {Swanger}}, \bibinfo {author} {\bibfnamefont {E.~D.}\ \bibnamefont {Rodriguez}}, \bibinfo {author} {\bibfnamefont {P.}~\bibnamefont {Bajada}}, \bibinfo {author} {\bibfnamefont {F.}~\bibnamefont {Jiao}}, \bibinfo {author} {\bibfnamefont {K.}~\bibnamefont {Peng}}, \bibinfo {author} {\bibfnamefont
  {A.}~\bibnamefont {Siahvashi}}, \bibinfo {author} {\bibfnamefont {M.~L.}\ \bibnamefont {Johns}},\ and\ \bibinfo {author} {\bibfnamefont {E.~F.}\ \bibnamefont {May}},\ }\bibfield  {title} {\bibinfo {title} {Hydrogen liquefaction: a review of the fundamental physics{,} engineering practice and future opportunities},\ }\href {https://doi.org/10.1039/D2EE00099G} {\bibfield  {journal} {\bibinfo  {journal} {Energy Environ. Sci.}\ }\textbf {\bibinfo {volume} {15}},\ \bibinfo {pages} {2690} (\bibinfo {year} {2022})}\BibitemShut {NoStop}%
\bibitem [{\citenamefont {Radenbaugh}(2004)}]{radenbaugh2004refrigeration}%
  \BibitemOpen
  \bibfield  {author} {\bibinfo {author} {\bibfnamefont {R.}~\bibnamefont {Radenbaugh}},\ }\bibfield  {title} {\bibinfo {title} {Refrigeration for superconductors},\ }\href@noop {} {\bibfield  {journal} {\bibinfo  {journal} {Proceedings of the IEEE}\ }\textbf {\bibinfo {volume} {92}},\ \bibinfo {pages} {1719} (\bibinfo {year} {2004})}\BibitemShut {NoStop}%
\bibitem [{\citenamefont {Farrah}\ \emph {et~al.}(2019)\citenamefont {Farrah}, \citenamefont {Smith}, \citenamefont {Ardila}, \citenamefont {Bradford}, \citenamefont {Dipirro}, \citenamefont {Ferkinhoff}, \citenamefont {Glenn}, \citenamefont {Goldsmith}, \citenamefont {Leisawitz}, \citenamefont {Nikola} \emph {et~al.}}]{farrah2019far}%
  \BibitemOpen
  \bibfield  {author} {\bibinfo {author} {\bibfnamefont {D.}~\bibnamefont {Farrah}}, \bibinfo {author} {\bibfnamefont {K.~E.}\ \bibnamefont {Smith}}, \bibinfo {author} {\bibfnamefont {D.}~\bibnamefont {Ardila}}, \bibinfo {author} {\bibfnamefont {C.~M.}\ \bibnamefont {Bradford}}, \bibinfo {author} {\bibfnamefont {M.}~\bibnamefont {Dipirro}}, \bibinfo {author} {\bibfnamefont {C.}~\bibnamefont {Ferkinhoff}}, \bibinfo {author} {\bibfnamefont {J.}~\bibnamefont {Glenn}}, \bibinfo {author} {\bibfnamefont {P.}~\bibnamefont {Goldsmith}}, \bibinfo {author} {\bibfnamefont {D.}~\bibnamefont {Leisawitz}}, \bibinfo {author} {\bibfnamefont {T.}~\bibnamefont {Nikola}}, \emph {et~al.},\ }\bibfield  {title} {\bibinfo {title} {Far-infrared instrumentation and technological development for the next decade},\ }\href@noop {} {\bibfield  {journal} {\bibinfo  {journal} {Journal of Astronomical Telescopes, Instruments, and Systems}\ }\textbf {\bibinfo {volume} {5}},\ \bibinfo {pages} {020901} (\bibinfo {year} {2019})}\BibitemShut
  {NoStop}%
\bibitem [{\citenamefont {Numazawa}\ \emph {et~al.}(2014)\citenamefont {Numazawa}, \citenamefont {Kamiya}, \citenamefont {Utaki},\ and\ \citenamefont {Matsumoto}}]{numazawa2014magnetic}%
  \BibitemOpen
  \bibfield  {author} {\bibinfo {author} {\bibfnamefont {T.}~\bibnamefont {Numazawa}}, \bibinfo {author} {\bibfnamefont {K.}~\bibnamefont {Kamiya}}, \bibinfo {author} {\bibfnamefont {T.}~\bibnamefont {Utaki}},\ and\ \bibinfo {author} {\bibfnamefont {K.}~\bibnamefont {Matsumoto}},\ }\bibfield  {title} {\bibinfo {title} {Magnetic refrigerator for hydrogen liquefaction},\ }\href@noop {} {\bibfield  {journal} {\bibinfo  {journal} {Cryogenics}\ }\textbf {\bibinfo {volume} {62}},\ \bibinfo {pages} {185} (\bibinfo {year} {2014})}\BibitemShut {NoStop}%
\bibitem [{\citenamefont {Ansarinasab}\ \emph {et~al.}(2023)\citenamefont {Ansarinasab}, \citenamefont {Fatimah},\ and\ \citenamefont {Khojasteh-Salkuyeh}}]{ansarinasab2023conceptual}%
  \BibitemOpen
  \bibfield  {author} {\bibinfo {author} {\bibfnamefont {H.}~\bibnamefont {Ansarinasab}}, \bibinfo {author} {\bibfnamefont {M.}~\bibnamefont {Fatimah}},\ and\ \bibinfo {author} {\bibfnamefont {Y.}~\bibnamefont {Khojasteh-Salkuyeh}},\ }\bibfield  {title} {\bibinfo {title} {Conceptual design of two novel hydrogen liquefaction processes using a multistage active magnetic refrigeration system},\ }\href@noop {} {\bibfield  {journal} {\bibinfo  {journal} {Applied Thermal Engineering}\ }\textbf {\bibinfo {volume} {230}},\ \bibinfo {pages} {120771} (\bibinfo {year} {2023})}\BibitemShut {NoStop}%
\bibitem [{\citenamefont {Feng}\ \emph {et~al.}(2020)\citenamefont {Feng}, \citenamefont {Chen},\ and\ \citenamefont {Ihnfeldt}}]{feng2020modeling}%
  \BibitemOpen
  \bibfield  {author} {\bibinfo {author} {\bibfnamefont {T.}~\bibnamefont {Feng}}, \bibinfo {author} {\bibfnamefont {R.}~\bibnamefont {Chen}},\ and\ \bibinfo {author} {\bibfnamefont {R.~V.}\ \bibnamefont {Ihnfeldt}},\ }\bibfield  {title} {\bibinfo {title} {Modeling of hydrogen liquefaction using magnetocaloric cycles with permanent magnets},\ }\href@noop {} {\bibfield  {journal} {\bibinfo  {journal} {International Journal of Refrigeration}\ }\textbf {\bibinfo {volume} {119}},\ \bibinfo {pages} {238} (\bibinfo {year} {2020})}\BibitemShut {NoStop}%
\bibitem [{\citenamefont {Kitanovski}(2020)}]{kitanovski2020energy}%
  \BibitemOpen
  \bibfield  {author} {\bibinfo {author} {\bibfnamefont {A.}~\bibnamefont {Kitanovski}},\ }\bibfield  {title} {\bibinfo {title} {Energy applications of magnetocaloric materials},\ }\href@noop {} {\bibfield  {journal} {\bibinfo  {journal} {Advanced Energy Materials}\ }\textbf {\bibinfo {volume} {10}},\ \bibinfo {pages} {1903741} (\bibinfo {year} {2020})}\BibitemShut {NoStop}%
\bibitem [{\citenamefont {Pecharsky}\ and\ \citenamefont {Gschneidner~Jr}(1999)}]{pecharsky1999magnetocaloric}%
  \BibitemOpen
  \bibfield  {author} {\bibinfo {author} {\bibfnamefont {V.~K.}\ \bibnamefont {Pecharsky}}\ and\ \bibinfo {author} {\bibfnamefont {K.~A.}\ \bibnamefont {Gschneidner~Jr}},\ }\bibfield  {title} {\bibinfo {title} {Magnetocaloric effect and magnetic refrigeration},\ }\href@noop {} {\bibfield  {journal} {\bibinfo  {journal} {Journal of Magnetism and Magnetic Materials}\ }\textbf {\bibinfo {volume} {200}},\ \bibinfo {pages} {44} (\bibinfo {year} {1999})}\BibitemShut {NoStop}%
\bibitem [{\citenamefont {Kumar}\ \emph {et~al.}(2024)\citenamefont {Kumar}, \citenamefont {Muhammad}, \citenamefont {Kim}, \citenamefont {Yi}, \citenamefont {Son},\ and\ \citenamefont {Oh}}]{kumar2024exploring}%
  \BibitemOpen
  \bibfield  {author} {\bibinfo {author} {\bibfnamefont {S.}~\bibnamefont {Kumar}}, \bibinfo {author} {\bibfnamefont {R.}~\bibnamefont {Muhammad}}, \bibinfo {author} {\bibfnamefont {S.}~\bibnamefont {Kim}}, \bibinfo {author} {\bibfnamefont {J.}~\bibnamefont {Yi}}, \bibinfo {author} {\bibfnamefont {K.}~\bibnamefont {Son}},\ and\ \bibinfo {author} {\bibfnamefont {H.}~\bibnamefont {Oh}},\ }\bibfield  {title} {\bibinfo {title} {Exploring magnetocaloric materials for sustainable refrigeration near hydrogen gas liquefaction temperature},\ }\href@noop {} {\bibfield  {journal} {\bibinfo  {journal} {Advanced Functional Materials}\ }\textbf {\bibinfo {volume} {34}},\ \bibinfo {pages} {2402513} (\bibinfo {year} {2024})}\BibitemShut {NoStop}%
\bibitem [{\citenamefont {Guillou}\ \emph {et~al.}(2014)\citenamefont {Guillou}, \citenamefont {Porcari}, \citenamefont {Yibole}, \citenamefont {van Dijk},\ and\ \citenamefont {Br{\"u}ck}}]{guillou2014taming}%
  \BibitemOpen
  \bibfield  {author} {\bibinfo {author} {\bibfnamefont {F.}~\bibnamefont {Guillou}}, \bibinfo {author} {\bibfnamefont {G.}~\bibnamefont {Porcari}}, \bibinfo {author} {\bibfnamefont {H.}~\bibnamefont {Yibole}}, \bibinfo {author} {\bibfnamefont {N.}~\bibnamefont {van Dijk}},\ and\ \bibinfo {author} {\bibfnamefont {E.}~\bibnamefont {Br{\"u}ck}},\ }\bibfield  {title} {\bibinfo {title} {Taming the first-order transition in giant magnetocaloric materials},\ }\href@noop {} {\bibfield  {journal} {\bibinfo  {journal} {Advanced Materials}\ }\textbf {\bibinfo {volume} {26}},\ \bibinfo {pages} {2671} (\bibinfo {year} {2014})}\BibitemShut {NoStop}%
\bibitem [{\citenamefont {Pecharsky}\ and\ \citenamefont {Gschneidner~Jr}(1997)}]{pecharsky1997giant}%
  \BibitemOpen
  \bibfield  {author} {\bibinfo {author} {\bibfnamefont {V.~K.}\ \bibnamefont {Pecharsky}}\ and\ \bibinfo {author} {\bibfnamefont {K.~A.}\ \bibnamefont {Gschneidner~Jr}},\ }\bibfield  {title} {\bibinfo {title} {Giant magnetocaloric effect in {G}d$_5$({S}i$_2${G}e$_2$)},\ }\href@noop {} {\bibfield  {journal} {\bibinfo  {journal} {Phys. Rev. Lett.}\ }\textbf {\bibinfo {volume} {78}},\ \bibinfo {pages} {4494} (\bibinfo {year} {1997})}\BibitemShut {NoStop}%
\bibitem [{\citenamefont {McMichael}\ \emph {et~al.}(1992)\citenamefont {McMichael}, \citenamefont {Shull}, \citenamefont {Swartzendruber}, \citenamefont {Bennett},\ and\ \citenamefont {Watson}}]{mcmichael1992magnetocaloric}%
  \BibitemOpen
  \bibfield  {author} {\bibinfo {author} {\bibfnamefont {R.}~\bibnamefont {McMichael}}, \bibinfo {author} {\bibfnamefont {R.}~\bibnamefont {Shull}}, \bibinfo {author} {\bibfnamefont {L.}~\bibnamefont {Swartzendruber}}, \bibinfo {author} {\bibfnamefont {L.}~\bibnamefont {Bennett}},\ and\ \bibinfo {author} {\bibfnamefont {R.}~\bibnamefont {Watson}},\ }\bibfield  {title} {\bibinfo {title} {Magnetocaloric effect in superparamagnets},\ }\href@noop {} {\bibfield  {journal} {\bibinfo  {journal} {Journal of Magnetism and Magnetic Materials}\ }\textbf {\bibinfo {volume} {111}},\ \bibinfo {pages} {29} (\bibinfo {year} {1992})}\BibitemShut {NoStop}%
\bibitem [{\citenamefont {Mauger}(1983)}]{mauger1983magnetic}%
  \BibitemOpen
  \bibfield  {author} {\bibinfo {author} {\bibfnamefont {A.}~\bibnamefont {Mauger}},\ }\bibfield  {title} {\bibinfo {title} {Magnetic polaron: Theory and experiment},\ }\href@noop {} {\bibfield  {journal} {\bibinfo  {journal} {Physical Review B}\ }\textbf {\bibinfo {volume} {27}},\ \bibinfo {pages} {2308} (\bibinfo {year} {1983})}\BibitemShut {NoStop}%
\bibitem [{\citenamefont {Teresa}\ \emph {et~al.}(1997)\citenamefont {Teresa}, \citenamefont {Ibarra}, \citenamefont {Algarabel}, \citenamefont {Ritter}, \citenamefont {Marquina}, \citenamefont {Blasco}, \citenamefont {Garcia}, \citenamefont {del Moral},\ and\ \citenamefont {Arnold}}]{teresa1997evidence}%
  \BibitemOpen
  \bibfield  {author} {\bibinfo {author} {\bibfnamefont {J.~d.}\ \bibnamefont {Teresa}}, \bibinfo {author} {\bibfnamefont {M.}~\bibnamefont {Ibarra}}, \bibinfo {author} {\bibfnamefont {P.}~\bibnamefont {Algarabel}}, \bibinfo {author} {\bibfnamefont {C.}~\bibnamefont {Ritter}}, \bibinfo {author} {\bibfnamefont {C.}~\bibnamefont {Marquina}}, \bibinfo {author} {\bibfnamefont {J.}~\bibnamefont {Blasco}}, \bibinfo {author} {\bibfnamefont {J.}~\bibnamefont {Garcia}}, \bibinfo {author} {\bibfnamefont {A.}~\bibnamefont {del Moral}},\ and\ \bibinfo {author} {\bibfnamefont {Z.}~\bibnamefont {Arnold}},\ }\bibfield  {title} {\bibinfo {title} {Evidence for magnetic polarons in the magnetoresistive perovskites},\ }\href@noop {} {\bibfield  {journal} {\bibinfo  {journal} {Nature}\ }\textbf {\bibinfo {volume} {386}},\ \bibinfo {pages} {256} (\bibinfo {year} {1997})}\BibitemShut {NoStop}%
\bibitem [{\citenamefont {Koepsell}\ \emph {et~al.}(2019)\citenamefont {Koepsell}, \citenamefont {Vijayan}, \citenamefont {Sompet}, \citenamefont {Grusdt}, \citenamefont {Hilker}, \citenamefont {Demler}, \citenamefont {Salomon}, \citenamefont {Bloch},\ and\ \citenamefont {Gross}}]{koepsell2019imaging}%
  \BibitemOpen
  \bibfield  {author} {\bibinfo {author} {\bibfnamefont {J.}~\bibnamefont {Koepsell}}, \bibinfo {author} {\bibfnamefont {J.}~\bibnamefont {Vijayan}}, \bibinfo {author} {\bibfnamefont {P.}~\bibnamefont {Sompet}}, \bibinfo {author} {\bibfnamefont {F.}~\bibnamefont {Grusdt}}, \bibinfo {author} {\bibfnamefont {T.~A.}\ \bibnamefont {Hilker}}, \bibinfo {author} {\bibfnamefont {E.}~\bibnamefont {Demler}}, \bibinfo {author} {\bibfnamefont {G.}~\bibnamefont {Salomon}}, \bibinfo {author} {\bibfnamefont {I.}~\bibnamefont {Bloch}},\ and\ \bibinfo {author} {\bibfnamefont {C.}~\bibnamefont {Gross}},\ }\bibfield  {title} {\bibinfo {title} {Imaging magnetic polarons in the doped {Fermi--Hubbard} model},\ }\href@noop {} {\bibfield  {journal} {\bibinfo  {journal} {Nature}\ }\textbf {\bibinfo {volume} {572}},\ \bibinfo {pages} {358} (\bibinfo {year} {2019})}\BibitemShut {NoStop}%
\bibitem [{\citenamefont {Maksimov}\ \emph {et~al.}(2000)\citenamefont {Maksimov}, \citenamefont {Bacher}, \citenamefont {McDonald}, \citenamefont {Kulakovskii}, \citenamefont {Forchel}, \citenamefont {Becker}, \citenamefont {Landwehr},\ and\ \citenamefont {Molenkamp}}]{maksimov2000magnetic}%
  \BibitemOpen
  \bibfield  {author} {\bibinfo {author} {\bibfnamefont {A.}~\bibnamefont {Maksimov}}, \bibinfo {author} {\bibfnamefont {G.}~\bibnamefont {Bacher}}, \bibinfo {author} {\bibfnamefont {A.}~\bibnamefont {McDonald}}, \bibinfo {author} {\bibfnamefont {V.}~\bibnamefont {Kulakovskii}}, \bibinfo {author} {\bibfnamefont {A.}~\bibnamefont {Forchel}}, \bibinfo {author} {\bibfnamefont {C.}~\bibnamefont {Becker}}, \bibinfo {author} {\bibfnamefont {G.}~\bibnamefont {Landwehr}},\ and\ \bibinfo {author} {\bibfnamefont {L.}~\bibnamefont {Molenkamp}},\ }\bibfield  {title} {\bibinfo {title} {Magnetic polarons in a single diluted magnetic semiconductor quantum dot},\ }\href@noop {} {\bibfield  {journal} {\bibinfo  {journal} {Physical Review B}\ }\textbf {\bibinfo {volume} {62}},\ \bibinfo {pages} {R7767} (\bibinfo {year} {2000})}\BibitemShut {NoStop}%
\bibitem [{\citenamefont {Durst}\ \emph {et~al.}(2002)\citenamefont {Durst}, \citenamefont {Bhatt},\ and\ \citenamefont {Wolff}}]{durst2002bound}%
  \BibitemOpen
  \bibfield  {author} {\bibinfo {author} {\bibfnamefont {A.~C.}\ \bibnamefont {Durst}}, \bibinfo {author} {\bibfnamefont {R.}~\bibnamefont {Bhatt}},\ and\ \bibinfo {author} {\bibfnamefont {P.}~\bibnamefont {Wolff}},\ }\bibfield  {title} {\bibinfo {title} {Bound magnetic polaron interactions in insulating doped diluted magnetic semiconductors},\ }\href@noop {} {\bibfield  {journal} {\bibinfo  {journal} {Physical Review B}\ }\textbf {\bibinfo {volume} {65}},\ \bibinfo {pages} {235205} (\bibinfo {year} {2002})}\BibitemShut {NoStop}%
\bibitem [{\citenamefont {Snow}\ \emph {et~al.}(2001)\citenamefont {Snow}, \citenamefont {Cooper}, \citenamefont {Young}, \citenamefont {Fisk}, \citenamefont {Comment},\ and\ \citenamefont {Ansermet}}]{snow2001magnetic}%
  \BibitemOpen
  \bibfield  {author} {\bibinfo {author} {\bibfnamefont {C.}~\bibnamefont {Snow}}, \bibinfo {author} {\bibfnamefont {S.}~\bibnamefont {Cooper}}, \bibinfo {author} {\bibfnamefont {D.}~\bibnamefont {Young}}, \bibinfo {author} {\bibfnamefont {Z.}~\bibnamefont {Fisk}}, \bibinfo {author} {\bibfnamefont {A.}~\bibnamefont {Comment}},\ and\ \bibinfo {author} {\bibfnamefont {J.-P.}\ \bibnamefont {Ansermet}},\ }\bibfield  {title} {\bibinfo {title} {Magnetic polarons and the metal-semiconductor transitions in {(Eu, La)B$_6$} and {EuO}: {R}aman scattering studies},\ }\href@noop {} {\bibfield  {journal} {\bibinfo  {journal} {Physical Review B}\ }\textbf {\bibinfo {volume} {64}},\ \bibinfo {pages} {174412} (\bibinfo {year} {2001})}\BibitemShut {NoStop}%
\bibitem [{\citenamefont {Brooks}\ \emph {et~al.}(2004)\citenamefont {Brooks}, \citenamefont {Lancaster}, \citenamefont {Blundell}, \citenamefont {Hayes}, \citenamefont {Pratt},\ and\ \citenamefont {Fisk}}]{brooks2004magnetic}%
  \BibitemOpen
  \bibfield  {author} {\bibinfo {author} {\bibfnamefont {M.}~\bibnamefont {Brooks}}, \bibinfo {author} {\bibfnamefont {T.}~\bibnamefont {Lancaster}}, \bibinfo {author} {\bibfnamefont {S.}~\bibnamefont {Blundell}}, \bibinfo {author} {\bibfnamefont {W.}~\bibnamefont {Hayes}}, \bibinfo {author} {\bibfnamefont {F.}~\bibnamefont {Pratt}},\ and\ \bibinfo {author} {\bibfnamefont {Z.}~\bibnamefont {Fisk}},\ }\bibfield  {title} {\bibinfo {title} {Magnetic phase separation in {EuB$_6$} detected by muon spin rotation},\ }\href@noop {} {\bibfield  {journal} {\bibinfo  {journal} {Physical Review B}\ }\textbf {\bibinfo {volume} {70}},\ \bibinfo {pages} {020401} (\bibinfo {year} {2004})}\BibitemShut {NoStop}%
\bibitem [{\citenamefont {Zhang}\ \emph {et~al.}(2009)\citenamefont {Zhang}, \citenamefont {Yu}, \citenamefont {von Moln{\'a}r}, \citenamefont {Fisk},\ and\ \citenamefont {Xiong}}]{zhang2009nonlinear}%
  \BibitemOpen
  \bibfield  {author} {\bibinfo {author} {\bibfnamefont {X.}~\bibnamefont {Zhang}}, \bibinfo {author} {\bibfnamefont {L.}~\bibnamefont {Yu}}, \bibinfo {author} {\bibfnamefont {S.}~\bibnamefont {von Moln{\'a}r}}, \bibinfo {author} {\bibfnamefont {Z.}~\bibnamefont {Fisk}},\ and\ \bibinfo {author} {\bibfnamefont {P.}~\bibnamefont {Xiong}},\ }\bibfield  {title} {\bibinfo {title} {Nonlinear {H}all effect as a signature of electronic phase separation in the semimetallic ferromagnet {EuB$_6$}},\ }\href@noop {} {\bibfield  {journal} {\bibinfo  {journal} {Physical Review Letters}\ }\textbf {\bibinfo {volume} {103}},\ \bibinfo {pages} {106602} (\bibinfo {year} {2009})}\BibitemShut {NoStop}%
\bibitem [{\citenamefont {Beaudin}\ \emph {et~al.}(2026)\citenamefont {Beaudin}, \citenamefont {D{\'e}silets-Benoit}, \citenamefont {Bianchi}, \citenamefont {Arnold}, \citenamefont {Samothrakitis}, \citenamefont {Leos}, \citenamefont {Gavilano}, \citenamefont {Kenzelmann}, \citenamefont {Stenning}, \citenamefont {Laver} \emph {et~al.}}]{beaudin2026observation}%
  \BibitemOpen
  \bibfield  {author} {\bibinfo {author} {\bibfnamefont {G.}~\bibnamefont {Beaudin}}, \bibinfo {author} {\bibfnamefont {A.}~\bibnamefont {D{\'e}silets-Benoit}}, \bibinfo {author} {\bibfnamefont {A.~D.}\ \bibnamefont {Bianchi}}, \bibinfo {author} {\bibfnamefont {R.}~\bibnamefont {Arnold}}, \bibinfo {author} {\bibfnamefont {S.}~\bibnamefont {Samothrakitis}}, \bibinfo {author} {\bibfnamefont {N.~A.~G.}\ \bibnamefont {Leos}}, \bibinfo {author} {\bibfnamefont {J.~L.}\ \bibnamefont {Gavilano}}, \bibinfo {author} {\bibfnamefont {M.}~\bibnamefont {Kenzelmann}}, \bibinfo {author} {\bibfnamefont {K.~D.}\ \bibnamefont {Stenning}}, \bibinfo {author} {\bibfnamefont {M.}~\bibnamefont {Laver}}, \emph {et~al.},\ }\bibfield  {title} {\bibinfo {title} {Observation of magnetic polarons in {EuB$_6$} by small-angle neutron scattering},\ }\href@noop {} {\bibfield  {journal} {\bibinfo  {journal} {Physical Review B}\ }\textbf {\bibinfo {volume} {113}},\ \bibinfo {pages} {115127} (\bibinfo {year} {2026})}\BibitemShut {NoStop}%
\bibitem [{\citenamefont {Guy}\ \emph {et~al.}(1980)\citenamefont {Guy}, \citenamefont {Von~Molnar}, \citenamefont {Etourneau},\ and\ \citenamefont {Fisk}}]{guy1980charge}%
  \BibitemOpen
  \bibfield  {author} {\bibinfo {author} {\bibfnamefont {C.}~\bibnamefont {Guy}}, \bibinfo {author} {\bibfnamefont {S.}~\bibnamefont {Von~Molnar}}, \bibinfo {author} {\bibfnamefont {J.}~\bibnamefont {Etourneau}},\ and\ \bibinfo {author} {\bibfnamefont {Z.}~\bibnamefont {Fisk}},\ }\bibfield  {title} {\bibinfo {title} {Charge transport and pressure dependence of {T$_c$} of single crystal, ferromagnetic {EuB$_6$}},\ }\href@noop {} {\bibfield  {journal} {\bibinfo  {journal} {Solid State Communications}\ }\textbf {\bibinfo {volume} {33}},\ \bibinfo {pages} {1055} (\bibinfo {year} {1980})}\BibitemShut {NoStop}%
\bibitem [{\citenamefont {Henggeler}\ \emph {et~al.}(1998)\citenamefont {Henggeler}, \citenamefont {Ott}, \citenamefont {Young},\ and\ \citenamefont {Fisk}}]{henggeler1998magnetic}%
  \BibitemOpen
  \bibfield  {author} {\bibinfo {author} {\bibfnamefont {W.}~\bibnamefont {Henggeler}}, \bibinfo {author} {\bibfnamefont {H.-R.}\ \bibnamefont {Ott}}, \bibinfo {author} {\bibfnamefont {D.}~\bibnamefont {Young}},\ and\ \bibinfo {author} {\bibfnamefont {Z.}~\bibnamefont {Fisk}},\ }\bibfield  {title} {\bibinfo {title} {Magnetic ordering in {EuB$_6$}, investigated by neutron diffraction},\ }\href@noop {} {\bibfield  {journal} {\bibinfo  {journal} {Solid State Communications}\ }\textbf {\bibinfo {volume} {108}},\ \bibinfo {pages} {929} (\bibinfo {year} {1998})}\BibitemShut {NoStop}%
\bibitem [{\citenamefont {S{\"u}llow}\ \emph {et~al.}(2000)\citenamefont {S{\"u}llow}, \citenamefont {Prasad}, \citenamefont {Aronson}, \citenamefont {Bogdanovich}, \citenamefont {Sarrao},\ and\ \citenamefont {Fisk}}]{sullow2000metallization}%
  \BibitemOpen
  \bibfield  {author} {\bibinfo {author} {\bibfnamefont {S.}~\bibnamefont {S{\"u}llow}}, \bibinfo {author} {\bibfnamefont {I.}~\bibnamefont {Prasad}}, \bibinfo {author} {\bibfnamefont {M.}~\bibnamefont {Aronson}}, \bibinfo {author} {\bibfnamefont {S.}~\bibnamefont {Bogdanovich}}, \bibinfo {author} {\bibfnamefont {J.}~\bibnamefont {Sarrao}},\ and\ \bibinfo {author} {\bibfnamefont {Z.}~\bibnamefont {Fisk}},\ }\bibfield  {title} {\bibinfo {title} {Metallization and magnetic order in {EuB$_6$}},\ }\href@noop {} {\bibfield  {journal} {\bibinfo  {journal} {Physical Review B}\ }\textbf {\bibinfo {volume} {62}},\ \bibinfo {pages} {11626} (\bibinfo {year} {2000})}\BibitemShut {NoStop}%
\bibitem [{\citenamefont {Castro}\ \emph {et~al.}(2020)\citenamefont {Castro}, \citenamefont {Terashima}, \citenamefont {Yamamoto}, \citenamefont {Hou}, \citenamefont {Iwasaki}, \citenamefont {Matsumoto}, \citenamefont {Adachi}, \citenamefont {Saito}, \citenamefont {Song}, \citenamefont {Takeya} \emph {et~al.}}]{castro2020machine}%
  \BibitemOpen
  \bibfield  {author} {\bibinfo {author} {\bibfnamefont {P.~B.~d.}\ \bibnamefont {Castro}}, \bibinfo {author} {\bibfnamefont {K.}~\bibnamefont {Terashima}}, \bibinfo {author} {\bibfnamefont {T.~D.}\ \bibnamefont {Yamamoto}}, \bibinfo {author} {\bibfnamefont {Z.}~\bibnamefont {Hou}}, \bibinfo {author} {\bibfnamefont {S.}~\bibnamefont {Iwasaki}}, \bibinfo {author} {\bibfnamefont {R.}~\bibnamefont {Matsumoto}}, \bibinfo {author} {\bibfnamefont {S.}~\bibnamefont {Adachi}}, \bibinfo {author} {\bibfnamefont {Y.}~\bibnamefont {Saito}}, \bibinfo {author} {\bibfnamefont {P.}~\bibnamefont {Song}}, \bibinfo {author} {\bibfnamefont {H.}~\bibnamefont {Takeya}}, \emph {et~al.},\ }\bibfield  {title} {\bibinfo {title} {Machine-learning-guided discovery of the gigantic magnetocaloric effect in {HoB$_2$} near the hydrogen liquefaction temperature},\ }\href@noop {} {\bibfield  {journal} {\bibinfo  {journal} {NPG Asia Materials}\ }\textbf {\bibinfo {volume} {12}},\ \bibinfo {pages} {35} (\bibinfo {year} {2020})}\BibitemShut
  {NoStop}%
\bibitem [{\citenamefont {Yamamoto}\ \emph {et~al.}(2004)\citenamefont {Yamamoto}, \citenamefont {Nakagawa}, \citenamefont {Sako}, \citenamefont {Arakawa},\ and\ \citenamefont {Nitani}}]{YAMAMOTO2003Entropy}%
  \BibitemOpen
  \bibfield  {author} {\bibinfo {author} {\bibfnamefont {T.~A.}\ \bibnamefont {Yamamoto}}, \bibinfo {author} {\bibfnamefont {T.}~\bibnamefont {Nakagawa}}, \bibinfo {author} {\bibfnamefont {K.}~\bibnamefont {Sako}}, \bibinfo {author} {\bibfnamefont {T.}~\bibnamefont {Arakawa}},\ and\ \bibinfo {author} {\bibfnamefont {H.}~\bibnamefont {Nitani}},\ }\bibfield  {title} {\bibinfo {title} {Magnetocaloric effect of rare earth mono-nitrides, {TbN} and {HoN}},\ }\href {https://doi.org/https://doi.org/10.1016/j.jallcom.2003.12.012} {\bibfield  {journal} {\bibinfo  {journal} {Journal of Alloys and Compounds}\ }\textbf {\bibinfo {volume} {376}},\ \bibinfo {pages} {17} (\bibinfo {year} {2004})}\BibitemShut {NoStop}%
\bibitem [{\citenamefont {Ibarra-Gayt{\'a}n}\ \emph {et~al.}(2013)\citenamefont {Ibarra-Gayt{\'a}n}, \citenamefont {S{\'a}nchez-Vald{\'e}s}, \citenamefont {S{\'a}nchez-Llamazares}, \citenamefont {{\'A}lvarez-Alonso}, \citenamefont {Gorr{\'i}a},\ and\ \citenamefont {Blanco}}]{IbarraGaytan2013Entropy}%
  \BibitemOpen
  \bibfield  {author} {\bibinfo {author} {\bibfnamefont {P.~J.}\ \bibnamefont {Ibarra-Gayt{\'a}n}}, \bibinfo {author} {\bibfnamefont {C.~F.}\ \bibnamefont {S{\'a}nchez-Vald{\'e}s}}, \bibinfo {author} {\bibfnamefont {J.~L.}\ \bibnamefont {S{\'a}nchez-Llamazares}}, \bibinfo {author} {\bibfnamefont {P.}~\bibnamefont {{\'A}lvarez-Alonso}}, \bibinfo {author} {\bibfnamefont {P.}~\bibnamefont {Gorr{\'i}a}},\ and\ \bibinfo {author} {\bibfnamefont {J.~A.}\ \bibnamefont {Blanco}},\ }\bibfield  {title} {\bibinfo {title} {Texture-induced enhancement of the magnetocaloric response in melt-spun {DyNi$_2$} ribbons},\ }\href {https://doi.org/10.1063/1.4824073} {\bibfield  {journal} {\bibinfo  {journal} {Applied Physics Letters}\ }\textbf {\bibinfo {volume} {103}},\ \bibinfo {pages} {152401} (\bibinfo {year} {2013})}\BibitemShut {NoStop}%
\bibitem [{\citenamefont {Dong}\ \emph {et~al.}(2015)\citenamefont {Dong}, \citenamefont {Ma}, \citenamefont {Ke}, \citenamefont {Zhang}, \citenamefont {Wang}, \citenamefont {Shen}, \citenamefont {Sun},\ and\ \citenamefont {Cheng}}]{DONG2015Entropy}%
  \BibitemOpen
  \bibfield  {author} {\bibinfo {author} {\bibfnamefont {Q.}~\bibnamefont {Dong}}, \bibinfo {author} {\bibfnamefont {Y.}~\bibnamefont {Ma}}, \bibinfo {author} {\bibfnamefont {Y.}~\bibnamefont {Ke}}, \bibinfo {author} {\bibfnamefont {X.}~\bibnamefont {Zhang}}, \bibinfo {author} {\bibfnamefont {L.}~\bibnamefont {Wang}}, \bibinfo {author} {\bibfnamefont {B.}~\bibnamefont {Shen}}, \bibinfo {author} {\bibfnamefont {J.}~\bibnamefont {Sun}},\ and\ \bibinfo {author} {\bibfnamefont {Z.}~\bibnamefont {Cheng}},\ }\bibfield  {title} {\bibinfo {title} {Ericsson-like giant magnetocaloric effect in {GdCrO$_4$–ErCrO$_4$} composite oxides near liquid hydrogen temperature},\ }\href {https://doi.org/https://doi.org/10.1016/j.matlet.2015.09.070} {\bibfield  {journal} {\bibinfo  {journal} {Materials Letters}\ }\textbf {\bibinfo {volume} {161}},\ \bibinfo {pages} {669} (\bibinfo {year} {2015})}\BibitemShut {NoStop}%
\bibitem [{\citenamefont {Ćwik}\ \emph {et~al.}(2017)\citenamefont {Ćwik}, \citenamefont {Koshkid’ko}, \citenamefont {{de Oliveira}}, \citenamefont {Nenkov}, \citenamefont {Hackemer}, \citenamefont {Dilmieva}, \citenamefont {Kolchugina}, \citenamefont {Nikitin},\ and\ \citenamefont {Rogacki}}]{CWIK2017Entropy}%
  \BibitemOpen
  \bibfield  {author} {\bibinfo {author} {\bibfnamefont {J.}~\bibnamefont {Ćwik}}, \bibinfo {author} {\bibfnamefont {Y.}~\bibnamefont {Koshkid’ko}}, \bibinfo {author} {\bibfnamefont {N.}~\bibnamefont {{de Oliveira}}}, \bibinfo {author} {\bibfnamefont {K.}~\bibnamefont {Nenkov}}, \bibinfo {author} {\bibfnamefont {A.}~\bibnamefont {Hackemer}}, \bibinfo {author} {\bibfnamefont {E.}~\bibnamefont {Dilmieva}}, \bibinfo {author} {\bibfnamefont {N.}~\bibnamefont {Kolchugina}}, \bibinfo {author} {\bibfnamefont {S.}~\bibnamefont {Nikitin}},\ and\ \bibinfo {author} {\bibfnamefont {K.}~\bibnamefont {Rogacki}},\ }\bibfield  {title} {\bibinfo {title} {Magnetocaloric effect in {L}aves-phase rare-earth compounds with the second-order magnetic phase transition: Estimation of the high-field properties},\ }\href {https://doi.org/https://doi.org/10.1016/j.actamat.2017.05.054} {\bibfield  {journal} {\bibinfo  {journal} {Acta Materialia}\ }\textbf {\bibinfo {volume} {133}},\ \bibinfo {pages} {230} (\bibinfo {year}
  {2017})}\BibitemShut {NoStop}%
\bibitem [{\citenamefont {Zhang}\ \emph {et~al.}(2012)\citenamefont {Zhang}, \citenamefont {Xu}, \citenamefont {Zheng}, \citenamefont {Shen}, \citenamefont {Hua}, \citenamefont {Sun},\ and\ \citenamefont {Shen}}]{ZHANG2012Tad}%
  \BibitemOpen
  \bibfield  {author} {\bibinfo {author} {\bibfnamefont {H.}~\bibnamefont {Zhang}}, \bibinfo {author} {\bibfnamefont {Z.}~\bibnamefont {Xu}}, \bibinfo {author} {\bibfnamefont {X.}~\bibnamefont {Zheng}}, \bibinfo {author} {\bibfnamefont {J.}~\bibnamefont {Shen}}, \bibinfo {author} {\bibfnamefont {F.}~\bibnamefont {Hua}}, \bibinfo {author} {\bibfnamefont {J.}~\bibnamefont {Sun}},\ and\ \bibinfo {author} {\bibfnamefont {B.}~\bibnamefont {Shen}},\ }\bibfield  {title} {\bibinfo {title} {Giant magnetic refrigerant capacity in {Ho$_3$Al$_2$} compound},\ }\href {https://doi.org/https://doi.org/10.1016/j.ssc.2012.04.004} {\bibfield  {journal} {\bibinfo  {journal} {Solid State Communications}\ }\textbf {\bibinfo {volume} {152}},\ \bibinfo {pages} {1127} (\bibinfo {year} {2012})}\BibitemShut {NoStop}%
\bibitem [{\citenamefont {Li}\ \emph {et~al.}(2011)\citenamefont {Li}, \citenamefont {Huo}, \citenamefont {Igawa},\ and\ \citenamefont {Nishimura}}]{LI2011Tad}%
  \BibitemOpen
  \bibfield  {author} {\bibinfo {author} {\bibfnamefont {L.}~\bibnamefont {Li}}, \bibinfo {author} {\bibfnamefont {D.}~\bibnamefont {Huo}}, \bibinfo {author} {\bibfnamefont {H.}~\bibnamefont {Igawa}},\ and\ \bibinfo {author} {\bibfnamefont {K.}~\bibnamefont {Nishimura}},\ }\bibfield  {title} {\bibinfo {title} {Large magnetocaloric effect in {TbCo$_3$B$_2$} compound},\ }\href {https://doi.org/https://doi.org/10.1016/j.jallcom.2010.10.043} {\bibfield  {journal} {\bibinfo  {journal} {Journal of Alloys and Compounds}\ }\textbf {\bibinfo {volume} {509}},\ \bibinfo {pages} {1796} (\bibinfo {year} {2011})}\BibitemShut {NoStop}%
\bibitem [{\citenamefont {Süllow}\ \emph {et~al.}(1998)\citenamefont {Süllow}, \citenamefont {Prasad}, \citenamefont {Aronson}, \citenamefont {Sarrao}, \citenamefont {Fisk}, \citenamefont {Hristova}, \citenamefont {Lacerda}, \citenamefont {Hundley}, \citenamefont {Vigliante},\ and\ \citenamefont {Gibbs}}]{Sullow.1998.Structure}%
  \BibitemOpen
  \bibfield  {author} {\bibinfo {author} {\bibfnamefont {S.}~\bibnamefont {Süllow}}, \bibinfo {author} {\bibfnamefont {I.}~\bibnamefont {Prasad}}, \bibinfo {author} {\bibfnamefont {M.~C.}\ \bibnamefont {Aronson}}, \bibinfo {author} {\bibfnamefont {J.~L.}\ \bibnamefont {Sarrao}}, \bibinfo {author} {\bibfnamefont {Z.}~\bibnamefont {Fisk}}, \bibinfo {author} {\bibfnamefont {D.}~\bibnamefont {Hristova}}, \bibinfo {author} {\bibfnamefont {A.~H.}\ \bibnamefont {Lacerda}}, \bibinfo {author} {\bibfnamefont {M.~F.}\ \bibnamefont {Hundley}}, \bibinfo {author} {\bibfnamefont {A.}~\bibnamefont {Vigliante}},\ and\ \bibinfo {author} {\bibfnamefont {D.}~\bibnamefont {Gibbs}},\ }\bibfield  {title} {\bibinfo {title} {{Structure and magnetic order of {EuB$_6$}}},\ }\href {https://doi.org/10.1103/physrevb.57.5860} {\bibfield  {journal} {\bibinfo  {journal} {Physical Review B}\ }\textbf {\bibinfo {volume} {57}},\ \bibinfo {pages} {5860} (\bibinfo {year} {1998})}\BibitemShut {NoStop}%
\bibitem [{\citenamefont {Gomez~Alvarado}\ \emph {et~al.}(2024)\citenamefont {Gomez~Alvarado}, \citenamefont {Zoghlin}, \citenamefont {Jackson}, \citenamefont {Kautzsch}, \citenamefont {Plumb}, \citenamefont {Aling}, \citenamefont {Capa~Salinas}, \citenamefont {Pokharel}, \citenamefont {Pang}, \citenamefont {Gomez} \emph {et~al.}}]{gomez2024advances}%
  \BibitemOpen
  \bibfield  {author} {\bibinfo {author} {\bibfnamefont {S.~J.}\ \bibnamefont {Gomez~Alvarado}}, \bibinfo {author} {\bibfnamefont {E.}~\bibnamefont {Zoghlin}}, \bibinfo {author} {\bibfnamefont {A.}~\bibnamefont {Jackson}}, \bibinfo {author} {\bibfnamefont {L.}~\bibnamefont {Kautzsch}}, \bibinfo {author} {\bibfnamefont {J.}~\bibnamefont {Plumb}}, \bibinfo {author} {\bibfnamefont {M.}~\bibnamefont {Aling}}, \bibinfo {author} {\bibfnamefont {A.~N.}\ \bibnamefont {Capa~Salinas}}, \bibinfo {author} {\bibfnamefont {G.}~\bibnamefont {Pokharel}}, \bibinfo {author} {\bibfnamefont {Y.}~\bibnamefont {Pang}}, \bibinfo {author} {\bibfnamefont {R.~M.}\ \bibnamefont {Gomez}}, \emph {et~al.},\ }\bibfield  {title} {\bibinfo {title} {Advances in high-pressure laser floating zone growth: {The Laser Optical Kristallmacher II (LOKII)}},\ }\href@noop {} {\bibfield  {journal} {\bibinfo  {journal} {Review of Scientific Instruments}\ }\textbf {\bibinfo {volume} {95}} (\bibinfo {year} {2024})}\BibitemShut {NoStop}%
\bibitem [{\citenamefont {Kasaya}\ \emph {et~al.}(1978)\citenamefont {Kasaya}, \citenamefont {Tarascon}, \citenamefont {Etourneau},\ and\ \citenamefont {Hagenmuller}}]{kasaya1978study}%
  \BibitemOpen
  \bibfield  {author} {\bibinfo {author} {\bibfnamefont {M.}~\bibnamefont {Kasaya}}, \bibinfo {author} {\bibfnamefont {J.-M.}\ \bibnamefont {Tarascon}}, \bibinfo {author} {\bibfnamefont {J.}~\bibnamefont {Etourneau}},\ and\ \bibinfo {author} {\bibfnamefont {P.}~\bibnamefont {Hagenmuller}},\ }\bibfield  {title} {\bibinfo {title} {Study of carbon-substituted {EuB$_6$}},\ }\href@noop {} {\bibfield  {journal} {\bibinfo  {journal} {Materials Research Bulletin}\ }\textbf {\bibinfo {volume} {13}},\ \bibinfo {pages} {751} (\bibinfo {year} {1978})}\BibitemShut {NoStop}%
\bibitem [{\citenamefont {Gao}\ \emph {et~al.}(2021)\citenamefont {Gao}, \citenamefont {Xu}, \citenamefont {Li}, \citenamefont {Yi}, \citenamefont {Nie}, \citenamefont {Rao}, \citenamefont {Wang}, \citenamefont {Hu}, \citenamefont {Chen}, \citenamefont {Fan} \emph {et~al.}}]{gao2021time}%
  \BibitemOpen
  \bibfield  {author} {\bibinfo {author} {\bibfnamefont {S.-Y.}\ \bibnamefont {Gao}}, \bibinfo {author} {\bibfnamefont {S.}~\bibnamefont {Xu}}, \bibinfo {author} {\bibfnamefont {H.}~\bibnamefont {Li}}, \bibinfo {author} {\bibfnamefont {C.-J.}\ \bibnamefont {Yi}}, \bibinfo {author} {\bibfnamefont {S.-M.}\ \bibnamefont {Nie}}, \bibinfo {author} {\bibfnamefont {Z.-C.}\ \bibnamefont {Rao}}, \bibinfo {author} {\bibfnamefont {H.}~\bibnamefont {Wang}}, \bibinfo {author} {\bibfnamefont {Q.-X.}\ \bibnamefont {Hu}}, \bibinfo {author} {\bibfnamefont {X.-Z.}\ \bibnamefont {Chen}}, \bibinfo {author} {\bibfnamefont {W.-H.}\ \bibnamefont {Fan}}, \emph {et~al.},\ }\bibfield  {title} {\bibinfo {title} {Time-reversal symmetry breaking driven topological phase transition in {EuB$_6$}},\ }\href@noop {} {\bibfield  {journal} {\bibinfo  {journal} {Physical Review X}\ }\textbf {\bibinfo {volume} {11}},\ \bibinfo {pages} {021016} (\bibinfo {year} {2021})}\BibitemShut {NoStop}%
\bibitem [{\citenamefont {Degiorgi}\ \emph {et~al.}(1997)\citenamefont {Degiorgi}, \citenamefont {Felder}, \citenamefont {Ott}, \citenamefont {Sarrao},\ and\ \citenamefont {Fisk}}]{degiorgi1997low}%
  \BibitemOpen
  \bibfield  {author} {\bibinfo {author} {\bibfnamefont {L.}~\bibnamefont {Degiorgi}}, \bibinfo {author} {\bibfnamefont {E.}~\bibnamefont {Felder}}, \bibinfo {author} {\bibfnamefont {H.}~\bibnamefont {Ott}}, \bibinfo {author} {\bibfnamefont {J.}~\bibnamefont {Sarrao}},\ and\ \bibinfo {author} {\bibfnamefont {Z.}~\bibnamefont {Fisk}},\ }\bibfield  {title} {\bibinfo {title} {Low-temperature anomalies and ferromagnetism of {EuB$_6$}},\ }\href@noop {} {\bibfield  {journal} {\bibinfo  {journal} {Physical Review Letters}\ }\textbf {\bibinfo {volume} {79}},\ \bibinfo {pages} {5134} (\bibinfo {year} {1997})}\BibitemShut {NoStop}%
\bibitem [{\citenamefont {Yuan}\ \emph {et~al.}(2024)\citenamefont {Yuan}, \citenamefont {Yang}, \citenamefont {Patra},\ and\ \citenamefont {Liao}}]{Yuan2024enhancing}%
  \BibitemOpen
  \bibfield  {author} {\bibinfo {author} {\bibfnamefont {J.}~\bibnamefont {Yuan}}, \bibinfo {author} {\bibfnamefont {R.}~\bibnamefont {Yang}}, \bibinfo {author} {\bibfnamefont {L.}~\bibnamefont {Patra}},\ and\ \bibinfo {author} {\bibfnamefont {B.}~\bibnamefont {Liao}},\ }\bibfield  {title} {\bibinfo {title} {Enhancing magnetocaloric material discovery: A machine learning approach using an autogenerated database by large language models},\ }\href {https://doi.org/10.1063/5.0206855} {\bibfield  {journal} {\bibinfo  {journal} {AIP Advances}\ }\textbf {\bibinfo {volume} {14}},\ \bibinfo {pages} {085125} (\bibinfo {year} {2024})}\BibitemShut {NoStop}%
\bibitem [{\citenamefont {Bansmann}\ \emph {et~al.}(2005)\citenamefont {Bansmann}, \citenamefont {Baker}, \citenamefont {Binns}, \citenamefont {Blackman}, \citenamefont {Bucher}, \citenamefont {Dorantes-D{\'a}vila}, \citenamefont {Dupuis}, \citenamefont {Favre}, \citenamefont {Kechrakos}, \citenamefont {Kleibert} \emph {et~al.}}]{bansmann2005magnetic}%
  \BibitemOpen
  \bibfield  {author} {\bibinfo {author} {\bibfnamefont {J.}~\bibnamefont {Bansmann}}, \bibinfo {author} {\bibfnamefont {S.}~\bibnamefont {Baker}}, \bibinfo {author} {\bibfnamefont {C.}~\bibnamefont {Binns}}, \bibinfo {author} {\bibfnamefont {J.}~\bibnamefont {Blackman}}, \bibinfo {author} {\bibfnamefont {J.-P.}\ \bibnamefont {Bucher}}, \bibinfo {author} {\bibfnamefont {J.}~\bibnamefont {Dorantes-D{\'a}vila}}, \bibinfo {author} {\bibfnamefont {V.}~\bibnamefont {Dupuis}}, \bibinfo {author} {\bibfnamefont {L.}~\bibnamefont {Favre}}, \bibinfo {author} {\bibfnamefont {D.}~\bibnamefont {Kechrakos}}, \bibinfo {author} {\bibfnamefont {A.}~\bibnamefont {Kleibert}}, \emph {et~al.},\ }\bibfield  {title} {\bibinfo {title} {Magnetic and structural properties of isolated and assembled clusters},\ }\href@noop {} {\bibfield  {journal} {\bibinfo  {journal} {Surface Science Reports}\ }\textbf {\bibinfo {volume} {56}},\ \bibinfo {pages} {189} (\bibinfo {year} {2005})}\BibitemShut {NoStop}%
\bibitem [{\citenamefont {Hohenberg}\ and\ \citenamefont {Halperin}(1977)}]{hohenberg1977theory}%
  \BibitemOpen
  \bibfield  {author} {\bibinfo {author} {\bibfnamefont {P.~C.}\ \bibnamefont {Hohenberg}}\ and\ \bibinfo {author} {\bibfnamefont {B.~I.}\ \bibnamefont {Halperin}},\ }\bibfield  {title} {\bibinfo {title} {Theory of dynamic critical phenomena},\ }\href@noop {} {\bibfield  {journal} {\bibinfo  {journal} {Reviews of Modern Physics}\ }\textbf {\bibinfo {volume} {49}},\ \bibinfo {pages} {435} (\bibinfo {year} {1977})}\BibitemShut {NoStop}%
\bibitem [{\citenamefont {Storchak}\ \emph {et~al.}(2010)\citenamefont {Storchak}, \citenamefont {Eshchenko}, \citenamefont {Morenzoni}, \citenamefont {Ingle}, \citenamefont {Heiss}, \citenamefont {Schwarzl}, \citenamefont {Springholz}, \citenamefont {Kallaher},\ and\ \citenamefont {von Moln{\'a}r}}]{storchak2010magnetic}%
  \BibitemOpen
  \bibfield  {author} {\bibinfo {author} {\bibfnamefont {V.~G.}\ \bibnamefont {Storchak}}, \bibinfo {author} {\bibfnamefont {D.~G.}\ \bibnamefont {Eshchenko}}, \bibinfo {author} {\bibfnamefont {E.}~\bibnamefont {Morenzoni}}, \bibinfo {author} {\bibfnamefont {N.}~\bibnamefont {Ingle}}, \bibinfo {author} {\bibfnamefont {W.}~\bibnamefont {Heiss}}, \bibinfo {author} {\bibfnamefont {T.}~\bibnamefont {Schwarzl}}, \bibinfo {author} {\bibfnamefont {G.}~\bibnamefont {Springholz}}, \bibinfo {author} {\bibfnamefont {R.}~\bibnamefont {Kallaher}},\ and\ \bibinfo {author} {\bibfnamefont {S.}~\bibnamefont {von Moln{\'a}r}},\ }\bibfield  {title} {\bibinfo {title} {Magnetic polarons in {Eu}-based films of magnetic semiconductors},\ }\href@noop {} {\bibfield  {journal} {\bibinfo  {journal} {Physical Review B}\ }\textbf {\bibinfo {volume} {81}},\ \bibinfo {pages} {153201} (\bibinfo {year} {2010})}\BibitemShut {NoStop}%
\bibitem [{\citenamefont {Emin}(2012)}]{emin2012magnetic}%
  \BibitemOpen
  \bibfield  {author} {\bibinfo {author} {\bibfnamefont {D.}~\bibnamefont {Emin}},\ }\bibfield  {title} {\bibinfo {title} {Magnetic polarons and colossal magnetoresistance},\ }in\ \href@noop {} {\emph {\bibinfo {booktitle} {Polarons}}}\ (\bibinfo  {publisher} {Cambridge University Press},\ \bibinfo {address} {Cambridge, UK},\ \bibinfo {year} {2012})\ pp.\ \bibinfo {pages} {65--72}\BibitemShut {NoStop}%
\bibitem [{\citenamefont {Moreo}\ \emph {et~al.}(2000)\citenamefont {Moreo}, \citenamefont {Mayr}, \citenamefont {Feiguin}, \citenamefont {Yunoki},\ and\ \citenamefont {Dagotto}}]{moreo2000giant}%
  \BibitemOpen
  \bibfield  {author} {\bibinfo {author} {\bibfnamefont {A.}~\bibnamefont {Moreo}}, \bibinfo {author} {\bibfnamefont {M.}~\bibnamefont {Mayr}}, \bibinfo {author} {\bibfnamefont {A.}~\bibnamefont {Feiguin}}, \bibinfo {author} {\bibfnamefont {S.}~\bibnamefont {Yunoki}},\ and\ \bibinfo {author} {\bibfnamefont {E.}~\bibnamefont {Dagotto}},\ }\bibfield  {title} {\bibinfo {title} {Giant cluster coexistence in doped manganites and other compounds},\ }\href@noop {} {\bibfield  {journal} {\bibinfo  {journal} {Physical Review Letters}\ }\textbf {\bibinfo {volume} {84}},\ \bibinfo {pages} {5568} (\bibinfo {year} {2000})}\BibitemShut {NoStop}%
\bibitem [{\citenamefont {Calder{\'o}n}\ \emph {et~al.}(2000)\citenamefont {Calder{\'o}n}, \citenamefont {Brey},\ and\ \citenamefont {Littlewood}}]{calderon2000stability}%
  \BibitemOpen
  \bibfield  {author} {\bibinfo {author} {\bibfnamefont {M.}~\bibnamefont {Calder{\'o}n}}, \bibinfo {author} {\bibfnamefont {L.}~\bibnamefont {Brey}},\ and\ \bibinfo {author} {\bibfnamefont {P.}~\bibnamefont {Littlewood}},\ }\bibfield  {title} {\bibinfo {title} {Stability and dynamics of free magnetic polarons},\ }\href@noop {} {\bibfield  {journal} {\bibinfo  {journal} {Physical Review B}\ }\textbf {\bibinfo {volume} {62}},\ \bibinfo {pages} {3368} (\bibinfo {year} {2000})}\BibitemShut {NoStop}%
\bibitem [{\citenamefont {Qian}\ \emph {et~al.}(2021)\citenamefont {Qian}, \citenamefont {Han}, \citenamefont {Zheng}, \citenamefont {Chen}, \citenamefont {Tyagi}, \citenamefont {Li}, \citenamefont {Du}, \citenamefont {Zheng}, \citenamefont {Huang}, \citenamefont {Zhang} \emph {et~al.}}]{qian2021high}%
  \BibitemOpen
  \bibfield  {author} {\bibinfo {author} {\bibfnamefont {X.}~\bibnamefont {Qian}}, \bibinfo {author} {\bibfnamefont {D.}~\bibnamefont {Han}}, \bibinfo {author} {\bibfnamefont {L.}~\bibnamefont {Zheng}}, \bibinfo {author} {\bibfnamefont {J.}~\bibnamefont {Chen}}, \bibinfo {author} {\bibfnamefont {M.}~\bibnamefont {Tyagi}}, \bibinfo {author} {\bibfnamefont {Q.}~\bibnamefont {Li}}, \bibinfo {author} {\bibfnamefont {F.}~\bibnamefont {Du}}, \bibinfo {author} {\bibfnamefont {S.}~\bibnamefont {Zheng}}, \bibinfo {author} {\bibfnamefont {X.}~\bibnamefont {Huang}}, \bibinfo {author} {\bibfnamefont {S.}~\bibnamefont {Zhang}}, \emph {et~al.},\ }\bibfield  {title} {\bibinfo {title} {High-entropy polymer produces a giant electrocaloric effect at low fields},\ }\href@noop {} {\bibfield  {journal} {\bibinfo  {journal} {Nature}\ }\textbf {\bibinfo {volume} {600}},\ \bibinfo {pages} {664} (\bibinfo {year} {2021})}\BibitemShut {NoStop}%
\end{thebibliography}%

\end{document}


\title{Supplementary Information for: Magnetic Polarons Enable Exceptional Magnetocaloric Response} 

\author{Joshua Ancheta}
\thanks{These authors contributed equally.}
\affiliation{Department of Mechanical Engineering, University of California, Santa Barbara, CA 93106, USA}

\author{Caeli Benyacko}
\thanks{These authors contributed equally.}
\affiliation{Materials Department, University of California, Santa Barbara, CA 93106, USA}

\author{Fanghao Zhang}
\thanks{These authors contributed equally.}
\affiliation{Department of Mechanical Engineering, University of California, Santa Barbara, CA 93106, USA}

\author{Stephen D. Wilson}
\email{stephendwilson@ucsb.edu}
\affiliation{Materials Department, University of California, Santa Barbara, CA 93106, USA}

\author{Bolin Liao}
\email{bliao@ucsb.edu} \affiliation{Department of Mechanical Engineering, University of California, Santa Barbara, CA 93106, USA}

\maketitle

\renewcommand{\thefigure}{S\arabic{figure}}
\renewcommand{\thetable}{S\Roman{table}}
\renewcommand{\theequation}{S\arabic{equation}}

\section*{Supplementary Methods}

\subsection{Synthesis}
Polycrystalline EuB$_6$ was synthesized following the reaction pathway \ce{Eu2O3 + 15B -> 2EuB6 + 3BO} using 99.99\% purity europium (III) oxide (Eu$_2$O$_3$) and 95\% purity of boron (B) amorphous powder, both of which are products from Millipore Sigma. 
The reactants were ground in a mortar and pestle until the mixture appeared homogeneous. 
To ensure a complete solid state reaction, the powder was then loaded into a 10\,mm diameter balloon and pressed using a cold isostatic press (CIP) at 50000\,psi. 
The starting mass of each rod was recorded before being wrapped in tantalum foil and placed into an alumina crucible. 
The pellet was placed in a high vacuum furnace and reacted at 1400\,$^{\circ}$C for 5~hours.
To ensure a complete solid state reaction and minimize volatile byproducts, this process was repeated until the pellets reached a stable mass. 

After producing phase pure powder of EuB$_6$, the powder was pressed into a 6\,mm diameter balloon to form a seed and feed rod for floating zone crystal growth.
The seed and feed rods were sintered under vacuum to ensure uniform density prior to the crystal growth.
Following sintering, the rods were then loaded in a high pressure laser floating zone furnace equipped with 7 IPG lasers capable of reaching maximum output powers of 300~W each for a combined capacity of 2100~W.
The chamber was pressurized to 800~psi of gettered ultra high-purity argon to suppress volatility during the reaction.
The crystal growth was stabilized at 270~W of individual laser power for a combined total of 1890~W.
The zone was then translated at 10 mm/hr with and counter-rotated at 10~rpm. 
The total grown boule measured about 6~mm in height from which high-quality millimeter-sized crystals could be mechanically extracted for measurements.

\subsection{Characterization}
\begin{figure}[!htb]
\includegraphics[width=0.5\textwidth]{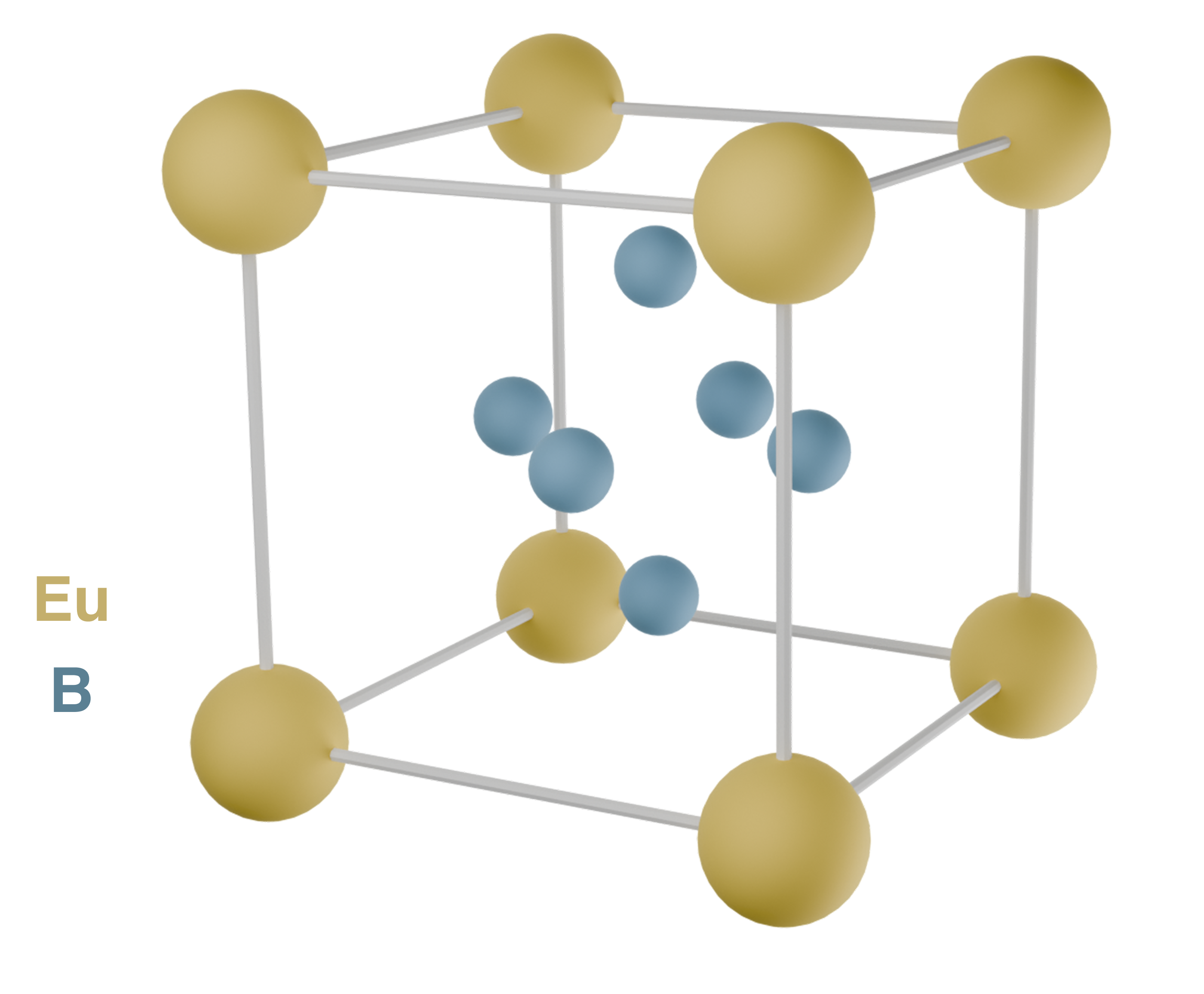}
\caption{\textbf{Crystal structure of EuB$_6$}. 
} 
\label{fig:structure}
\end{figure}

The crystal structure of EuB$_6$ is shown in Fig.~\ref{fig:structure}. The phase of the growth was confirmed via powder X-ray diffraction with a PANlytical Empyrean 3 X-ray diffractometer utilizing a Cu-source. 
Further, the orientation of the single crystals extracted from the as-grown boule were confirmed via Laue diffraction.

A Quantum Design Magnetic Properties Measurement System 3 (MPMS3) by Quantum Design to measure the magnetic moment as a function of temperature and applied magnetic field. Samples were fixed onto a quartz paddle using GE-varnish and measured using the vibrating sample magnetometry (VSM) option. To map out $\Delta S_m$, a sequence was used to measure the magnetic moment of the sample as the magnetic field increases in small intervals from 0~T to 5~T at each interval of temperature from 2~K to 30~K. After the measurements,  $\Delta S_m$ is extracted using Eq.~1 in the main text~\cite{bocarsly2018magnetoentropic}.

To measure the heat capacity (C$_p$) of EuB$_6$ as a function of temperature and external magnetic field, we used a Physical Properties Measurement System (PPMS) by Quantum Design. Thermal N-grease produced by Quantum Design was measured as the background addenda to 70~K at 0~T. Then the same sample measured previously in the MPMS3 was placed onto the puck with the (100) direction facing upward. The sequence was coded to start at 70~K cooling down to 30~K in 5~K intervals, then at 0.2~K intervals below 30~K. A single crystalline grain with a mass of 166.4 $\mu$g was used for the magnetization and specific heat measurements.

\begin{figure}[!htb]
\includegraphics[width=\textwidth]{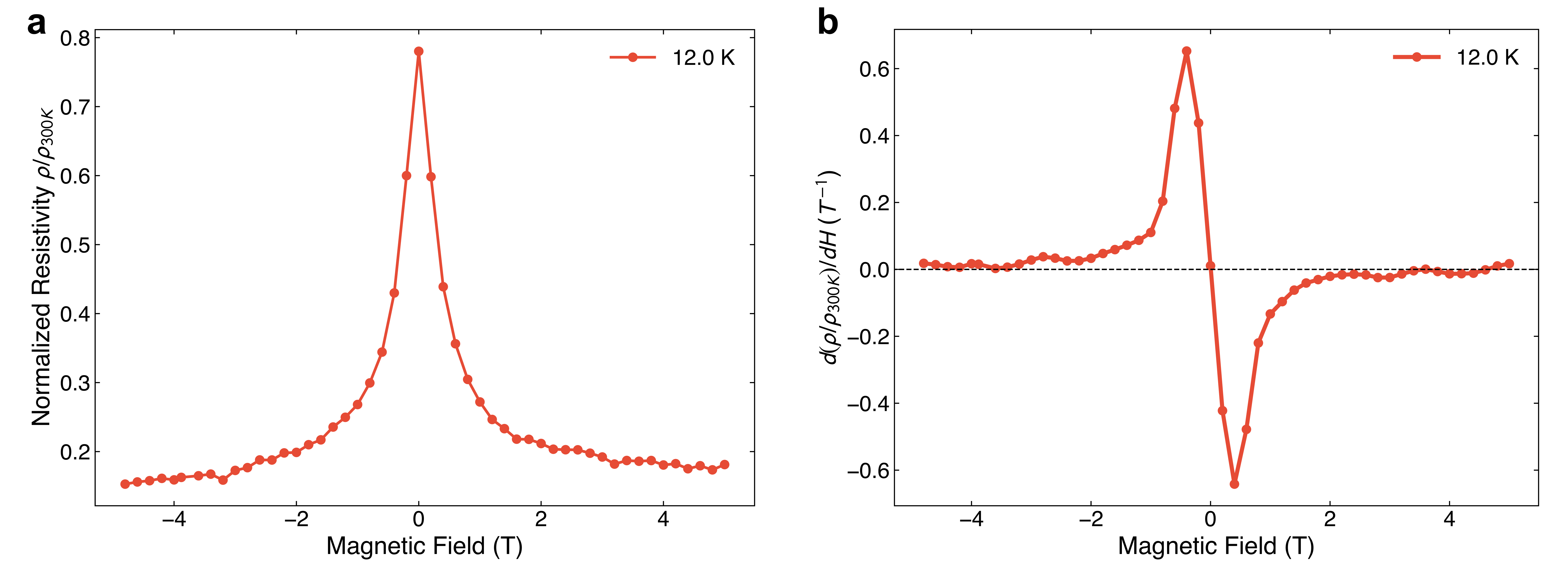}
\caption{\textbf{Magnetoresistance of EuB$_6$ at 12 K.}  \textbf{a,}  Field-dependent normalized resistivity $\rho(H)/\rho_{300K}$ at 12 K.
\textbf{b,} Differential magnetoresistance $d(\rho/\rho_{300K})/dH$ as a function of magnetic field. 
} 
\label{fig:MR_12K}
\end{figure}

The resistivity and magnetoresistance measurements were conducted on a polycrystalline sample  using the Electric Transport Option (ETO) of the PPMS. The temperature scan was performed from 300 K to 2 K. The temperature-dependent resistivity is presented in a normalized form, $\rho(T) / \rho(300\,\mathrm{K})$. The magnetoresistance was first performed at 12 K to check the symmetric MR in our sample, as shown in Fig.~\ref{fig:MR_12K}. The magnetoresistance sequence was coded to scan from 0 to 5 T in 0.1 T intervals. To minimize experimental noise in the derivative calculations, the raw magnetoresistance data were smoothed using a Savitzky-Golay filter with a window length of 7 points and a 3rd-order polynomial fit prior to differentiation.

\subsection{Direct measurement of the adiabatic temperature change ($\Delta T_{\mathrm{ad}}$)}

\begin{figure}[!htb]
\includegraphics[width=0.6\textwidth]{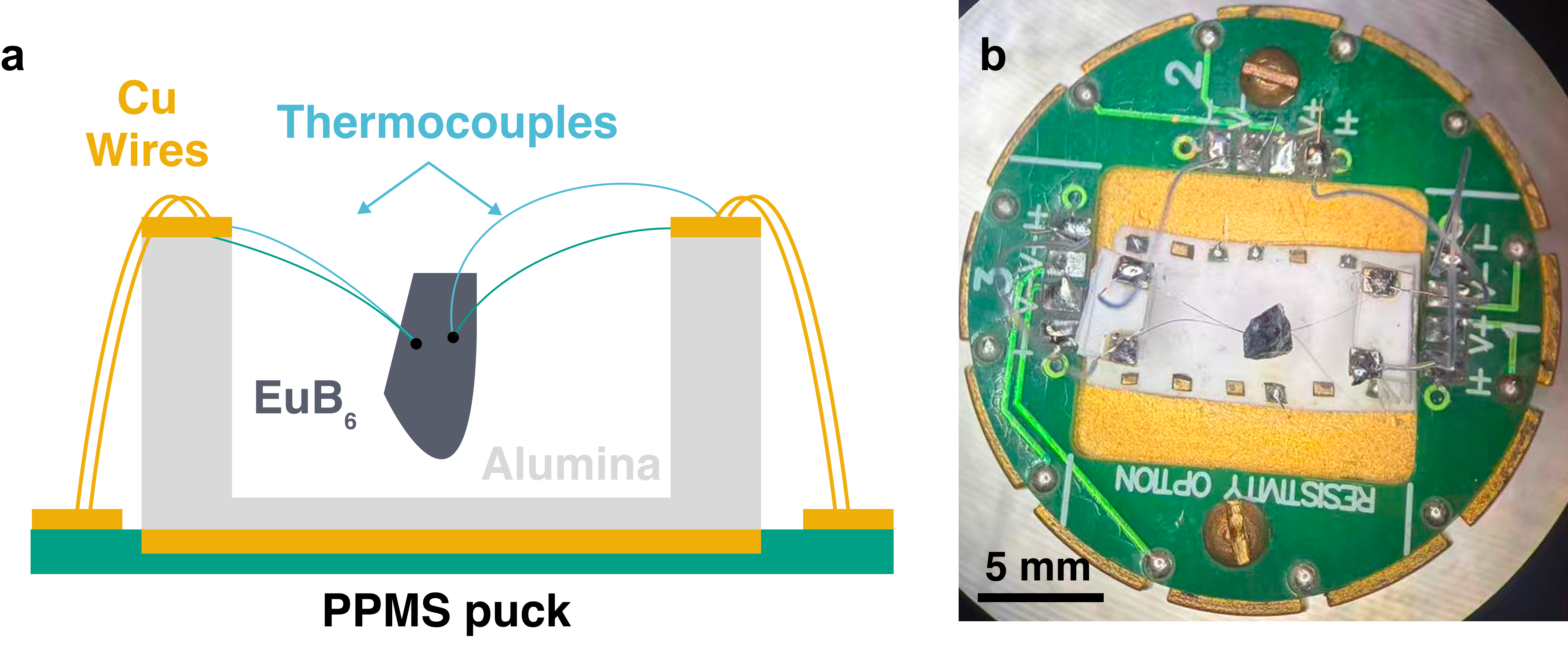}
\caption{\textbf{The setup of the direct measurement of the adiabatic temperature change.} The schematic [front view] (a) and digital photo [top view] (b) of the measurement device of the adiabatic temperature change.
} 
\label{fig:MCE_setup}
\end{figure}

 The adiabatic temperature change $\Delta T_{\mathrm{ad}}$ of EuB$_6$ was directly measured under a magnetic field ramp from 0 to 5~T at the $T_{\mathrm{c}}$($\sim$12~K) in PPMS chamber. 
The setup of the measurement is shown in the Fig.~\ref{fig:MCE_setup}.
 A large polycrystalline sample with a mass of 15 mg was used for this measurement. 
Specimen temperature was measured using two 25-$\mu m$ Type E thermocouples (Chromel/Constantan) while maintaining the sample in a suspended, high-vacuum configuration (10$^{-6}$ Torr) to mimic adiabatic conditions and reduce external thermal coupling.
The thermocouple junctions were encapsulated with Stycast to ensure robust thermal anchoring and negligible interfacial thermal resistance between the sensors and the specimen.
 The sample stage was attached on a PPMS resistivity puck using thermally conductive adhesive to ensure a good thermal and mechanical contact.

\begin{figure}[!htb]
\includegraphics[width=0.6\textwidth]{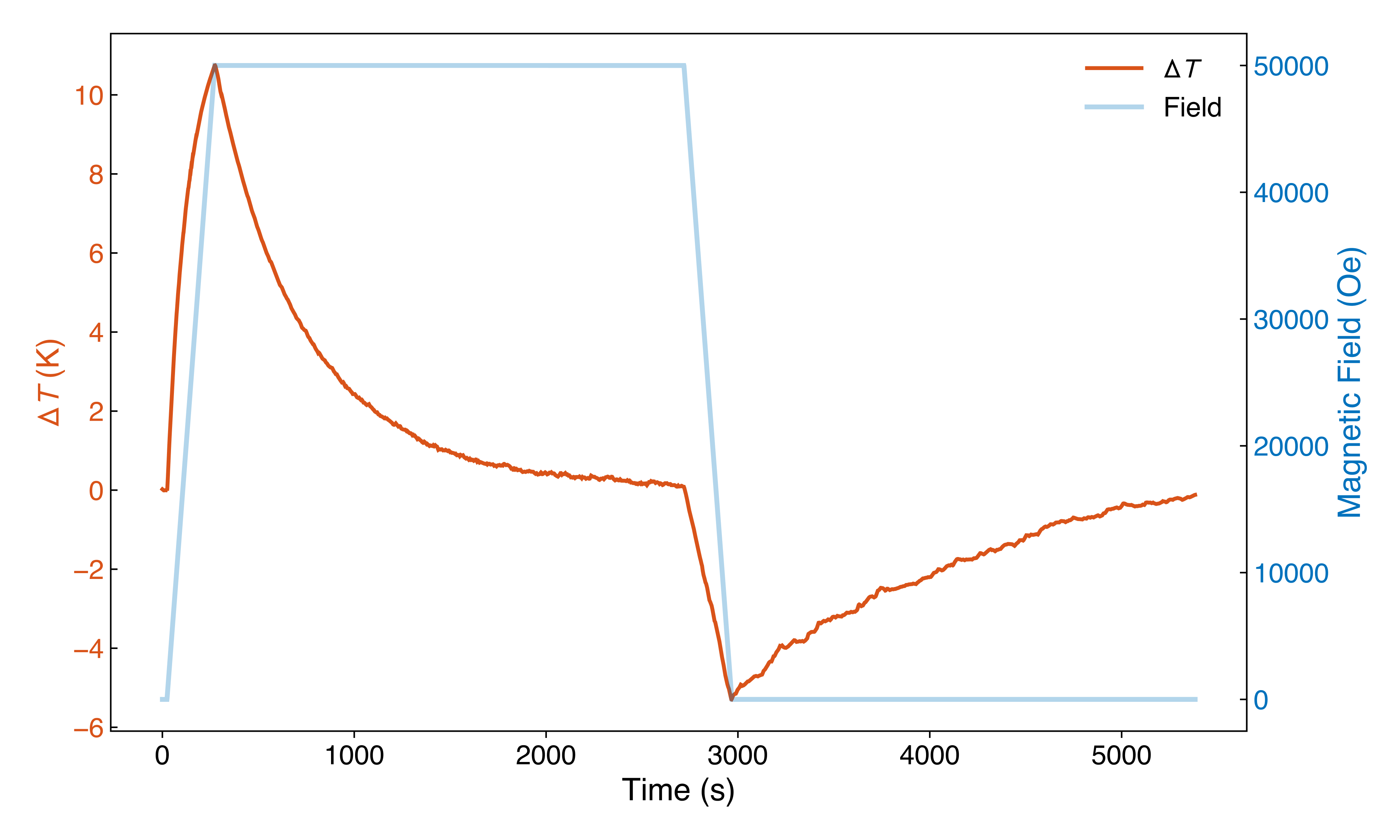}
\caption{\textbf{Direct measurement of the adiabatic temperature change.} The sample warmed up to large $\Delta T$ of 11 K when the field ramped up to 5 T, then cooled down with a lower $\Delta T$  of -5 K when the field decreased to 0 T.
} 
\label{fig:MCE_cooling}
\end{figure}

 The PPMS chamber was purged with helium before the measurement to ensure a stable base temperature. 
 The chamber was then pumped down to the high vacuum mode for the measurement.
 The voltage across the thermal couples were measured using the nanovoltmeter (Keithley, Model 2182A), with a 10$^{-8}$ V residue at 12 K. 
 The temperature differences were determined by directly converting the measured voltages using the Type E thermocouples' standardized Seebeck polynomials from the NIST ITS-90 database~\cite{burns1993temperature}.
 The magnetic field dependence of the Type E thermocouple sensitivity is minimal in this temperature range ($<$3\% at 5 T,~\cite{sample1974low}) and was treated as a systematic uncertainty.
 The magnetic field was ramped up from 0 to 5 T with a rate of 200 Oe/s, then held at 5 T until the voltage reading returned to the residue level, then the field was scanned from 5 T to 0 T with the same rate. The full warm-up and cool-down scan is shown here in Fig~\ref{fig:MCE_cooling}.

\subsection{Use of Large Language Model}
The authors acknowledge the use of ChatGPT (OpenAI, GPT-5.3) and Gemini (Google) as assistive tools in the preparation of this manuscript. The models were used to support drafting and editing of text, improving clarity and organization, and facilitating literature-oriented discussions. All scientific content, data analysis, interpretations, and conclusions were developed, verified, and validated by the authors. The authors take full responsibility for the accuracy and integrity of the work.
\clearpage

\section*{Supplementary Note 1: Background subtraction of phonon and electronic contributions in specific heat}

\begin{figure}[!htb]
\includegraphics[width=0.6\textwidth]{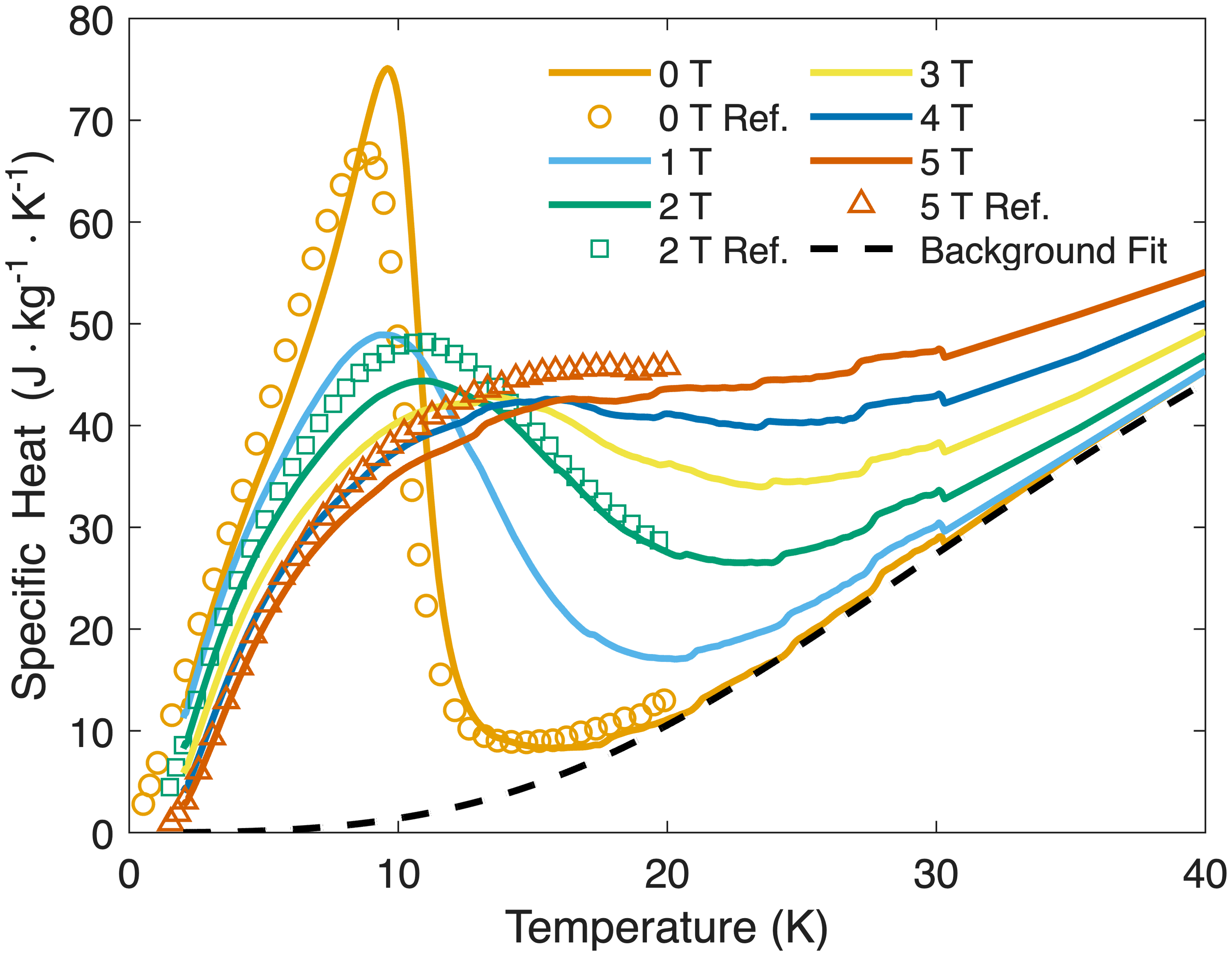}
\caption{\textbf{Measured specific heat of a single-crystalline grain of EuB$_6$ as a function of temperature and magnetic field.} Data represented by markers are from a previous report~\cite{Sullow.1998.Structure} for comparison. The background fit was used to isolate the magnetic contribution to the specific heat.
} 
\label{fig:cp}
\end{figure}

The measured specific heat of a single-crystalline grain of EuB$_6$ is shown here in Fig.~\ref{fig:cp}. Our measured data is compared to previously reported values below 20 K~\cite{Sullow.1998.Structure}.

To isolate the magnetic contribution to the specific heat, the measured total specific heat $C_p(T,H)$ is decomposed into electronic, phonon, and magnetic components:
\begin{equation}
C_p(T,H) = C_{\mathrm{el}}(T) + C_{\mathrm{ph}}(T) + C_{\mathrm{mag}}(T,H).
\end{equation}
The electronic and phonon contributions are assumed to be weakly dependent on magnetic field and are therefore treated as a common background for all field-dependent datasets.
The electronic contribution is described by a linear term,
\begin{equation}
C_{\mathrm{el}}(T) = \gamma T,
\end{equation}
where $\gamma$ is the electronic specific heat coefficient. The phonon contribution is modeled by explicitly accounting for both acoustic and optical modes. The acoustic phonons are described using a Debye model,
\begin{equation}
C_{\mathrm{D}}(T) = 9 N k_B \left( \frac{T}{\Theta_D} \right)^3 
\int_0^{\Theta_D/T} \frac{x^4 e^x}{(e^x - 1)^2} dx,
\end{equation}
where $\Theta_D$ is the Debye temperature and $N$ is the number of atoms per formula unit. To capture optical phonon contributions, one or more Einstein modes are included:
\begin{equation}
C_{\mathrm{E}}(T) = \sum_i 3 N_i k_B \left( \frac{\Theta_{E,i}}{T} \right)^2 
\frac{e^{\Theta_{E,i}/T}}{(e^{\Theta_{E,i}/T} - 1)^2},
\end{equation}
where $\Theta_{E,i}$ and $N_i$ are the characteristic temperature and mode weight of the $i$-th optical branch. Only one optical mode is used in the fitting in this work. The total phonon contribution is then given by
\begin{equation}
C_{\mathrm{ph}}(T) = C_{\mathrm{D}}(T) + C_{\mathrm{E}}(T).
\end{equation}

The background contributions $C_{\mathrm{el}}(T)$ and $C_{\mathrm{ph}}(T)$ are determined by fitting the total specific heat in a temperature range well above the magnetic transition, where the magnetic contribution is negligible or weakly varying. The fitting is performed simultaneously across multiple magnetic fields to ensure a consistent field-independent background. The fitting parameters include:
the electronic coefficient $\gamma$, the Debye temperature $\Theta_D$, and the Einstein temperature $\Theta_{E}$. The fitted background is shown as the black dashed line in Fig.~\ref{fig:cp}. The fitted electronic coefficient $\gamma$ is found to be negligibly small within the fitting uncertainty, indicating a very weak electronic contribution to the specific heat in EuB$_6$. This is consistent with the low carrier density and semimetallic character of the material. The phonon contribution is well captured by the combined Debye and Einstein model with a fitted $\Theta_D$ of 177.7 K and $\Theta_E$ of 931.3 K. The resulting background provides a consistent description across all measured magnetic fields. After subtraction of the electronic and phonon background, the remaining specific heat is dominated by the magnetic contribution, which exhibit strong field dependence and gives rise to the large magnetic entropy change discussed in the main text.

We note that this decomposition assumes that the phonon and electronic contributions are independent of magnetic field and that the phonon spectrum can be adequately described by a small number of Debye and Einstein modes. While these approximations are standard, small systematic uncertainties may remain, particularly at intermediate temperatures where magnetic and phonon contributions overlap. Nevertheless, the overall conclusions regarding the magnetic entropy and magnetocaloric response are robust against reasonable variations of the background model.
\clearpage

\section*{Supplementary Note 2: Low Temperature Extrapolation of Specific Heat}
\begin{figure}[htbp]
    \centering
   \includegraphics[width=0.6\textwidth]{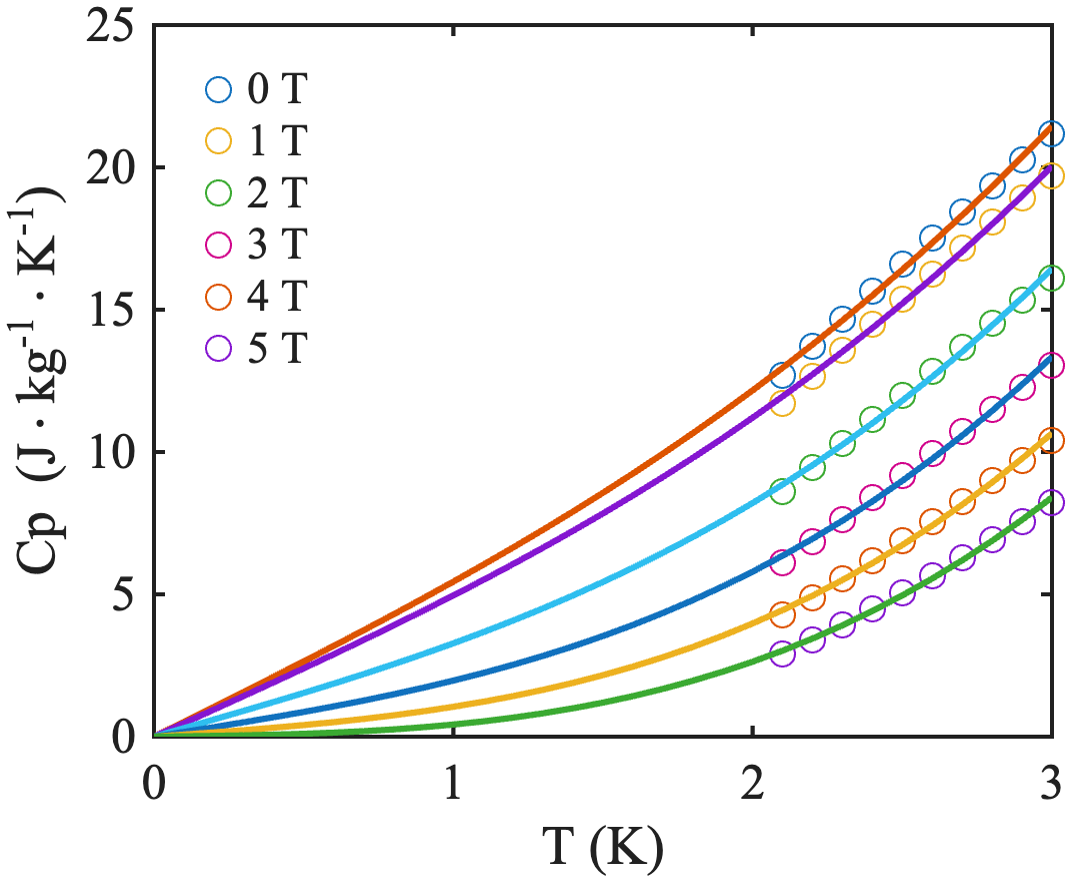}
    \caption{\textbf{Low temperature extrapolation of measured specific heat.} The circles are measured specific heat of a single crystal EuB$_6$ and the solid lines are polynomial fits.}
    \label{fig:fit}
    
\end{figure}

The lowest accessible temperature for our specific heat measurements is 2 K. To compute the total entropy $S(T,H)$ with Eqn. 1 in the main text, we need to extrapolate the measured specific heat down to 0 K to include the entropy contribution between 0 K and 2 K. For this purpose, we fit the specific heat data between 2 K and 3 K under each magnetic field to a polynomial form $C_p(T)=\gamma T+\beta T^3$, where the linear and cubic terms account for electronic and phononic contributions, respectively. The fitted results are shown in Fig.~\ref{fig:fit}. We also tried adding a magnetic term $\alpha T^{1.5}$ but then the fitting becomes unstable. 
\clearpage

\section*{Supplementary Note 3: Magnetocaloric Properties of E\MakeLowercase{u}B$_6$ Powders}

\begin{figure}[htbp]
    \centering
   \includegraphics[width=\textwidth]{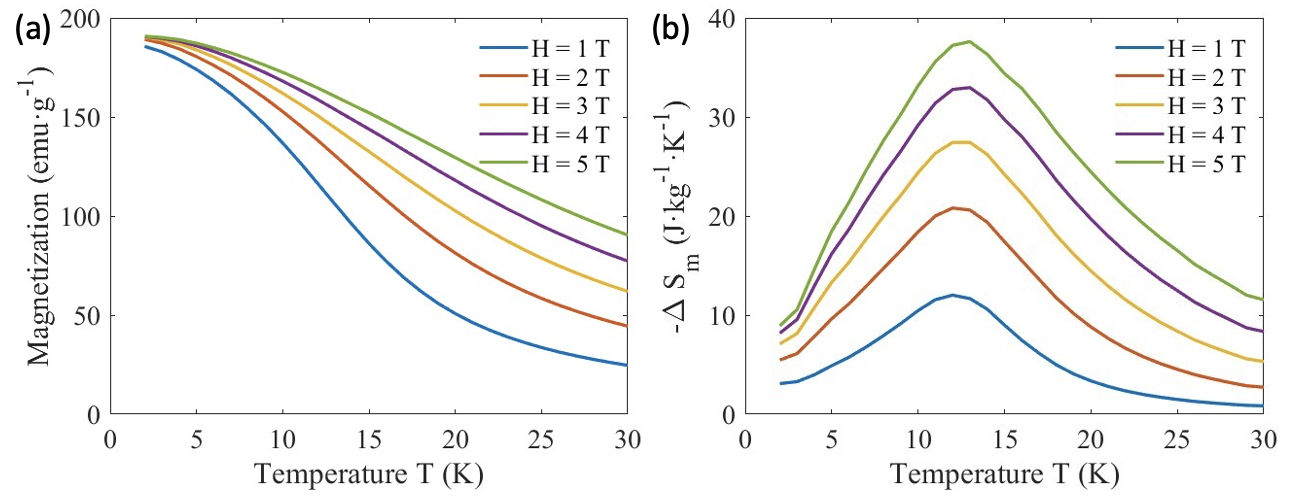}
    \caption{\textbf{Isothermal entropy change $\Delta S_M$ of EuB$_6$ powder.} (a) The magnetization of EuB$_6$ powder as a function of temperature measured with different applied magnetic fields. (c) Calculated isothermal entropy change $\Delta S_M$ of EuB$_6$ powder as a function of temperature and applied magnetic fields.}
    
\end{figure}

\begin{figure}[htbp]
    \centering
   \includegraphics[width=\textwidth]{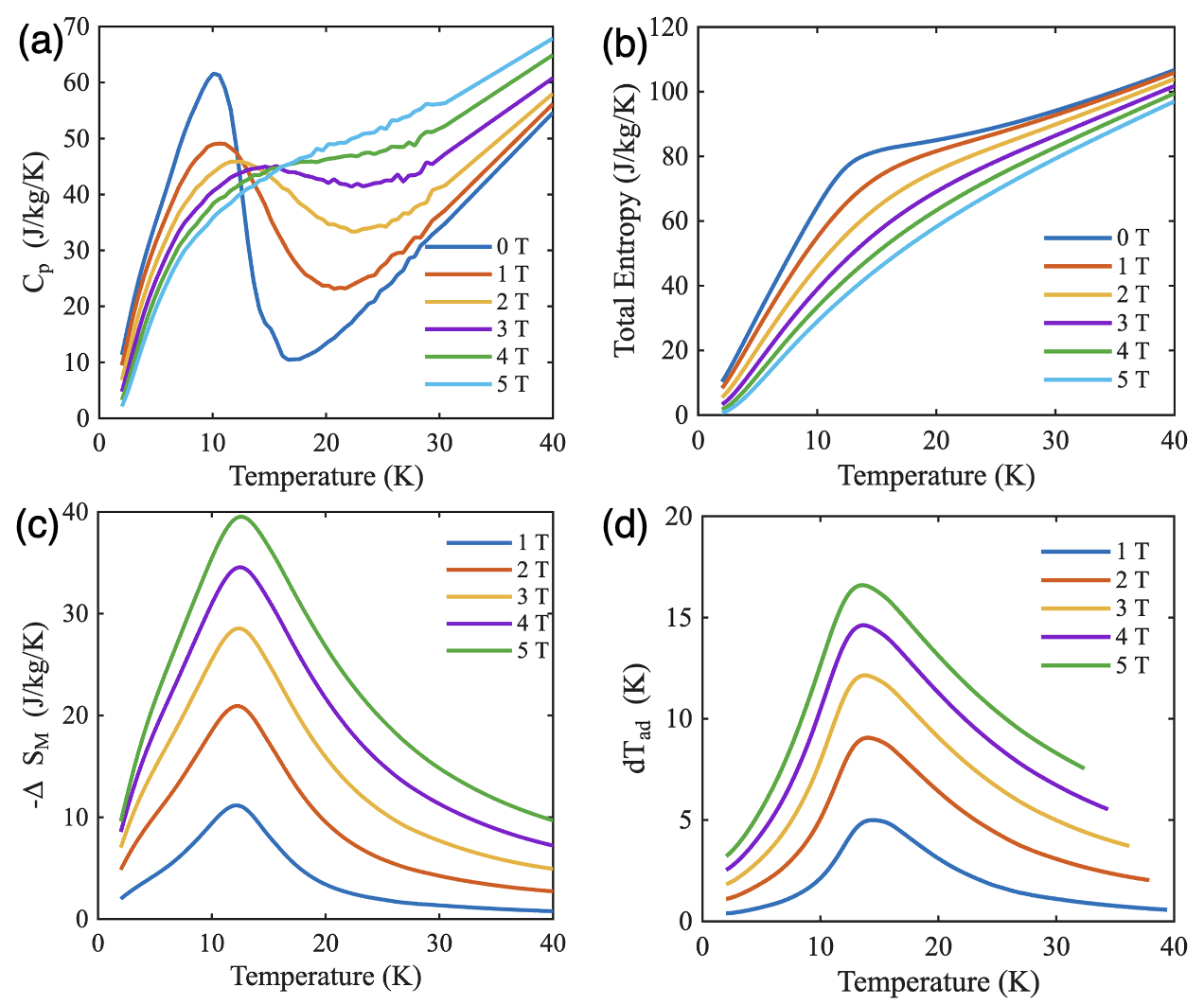}
    \caption{\textbf{Specific heat and total entropy of EuB$_6$ powder.} (a) Measured specific heat of EuB$_6$ powder as a function of temperature and magnetic field. (b) Calculated total entropy of EnB$_6$ powder as a function of temperature and magnetic field. (c) Calculated isothermal entropy change $\Delta S_{M}$ of EuB$_6$ powder as a function of temperature and magnetic field from the specific heat measurement. (d) Calculated adiabatic temperature $\Delta T_{ad}$ of EuB$_6$ powder as a function of temperature and magnetic field from the specific heat measurement.}
    
\end{figure}
\clearpage

\section*{Supplementary Note 4: Thermal model for direct measurement of the adiabatic temperature change}

To quantify the adiabatic temperature change $\Delta T_{\mathrm{ad}}$ from the direct temperature measurements, we model the transient thermal response of the sample using a minimal phenomenological framework~\cite{Franco2012magnetocaloric}. The goal of this model is to capture the essential features of the temperature evolution during magnetic field ramping while avoiding overparameterization.

The measured temperature $T(t)$ reflects a balance between (i) magnetocaloric heating associated with changes in magnetic field and (ii) thermal relaxation to the environment. We model the temperature evolution using a first-order differential equation,
\begin{equation}
\frac{dT}{dt} = \frac{\Delta T_{\mathrm{ad}}}{\tau_{\mathrm{ramp}}} \, g(t) - \frac{T - T_0}{\tau},
\end{equation}
where $T_0$ is the initial temperature, $\tau$ is an effective thermal relaxation time constant, and $\tau_{\mathrm{ramp}}$ characterizes the duration of the magnetic field ramp. The function $g(t)$ describes the time dependence of the magnetic field and is taken to be constant during the field ramp and zero otherwise, corresponding to a constant heating rate during field application. Under this approximation, the magnetocaloric contribution is modeled as a constant heating term during the ramp-up of the magnetic field, producing a total temperature increase of $\Delta T_{\mathrm{ad}}$ in the absence of thermal losses. After the field reaches its maximum value, the heating term vanishes and the system relaxes exponentially toward equilibrium.

The model reduces to two regimes: 1. Field ramp: During the field ramp, the temperature increases approximately linearly with time, with a slope determined by $\Delta T_{\mathrm{ad}}$, while simultaneously experiencing thermal relaxation. 2. Post-ramp relaxation: After the field reaches its maximum value, the temperature decays exponentially with a characteristic time $\tau$,
    \begin{equation}
    T(t) = T_{\mathrm{max}} \exp\left(-\frac{t - t_{\mathrm{ramp}}}{\tau}\right) + T_0,
    \end{equation}
where $T_{\mathrm{max}}$ is the peak temperature reached at the end of the ramp. The full time-dependent temperature trace is fitted using nonlinear least-squares regression. The only free parameters in the model are the adiabatic temperature change $\Delta T_{\mathrm{ad}}$ and the relaxation time constant $\tau$. The initial temperature $T_0$ and the ramp duration are fixed by the experimental data. A comparison between the experimental data and the model fitting is provided in Fig.~3d in the Main Text.

Using this model, we obtain an effective adiabatic temperature change of $\Delta T_{\mathrm{ad}} \approx 15.1$~K under a 5~T field and a thermal relaxation time of about 430 s. This $\Delta T_{\mathrm{ad}}$ value is in good agreement with the magnitude expected from thermodynamic analysis of the powder sample based on specific heat, taking into account the non-ideal experimental conditions and finite thermal coupling to the environment.

This model assumes a constant heating rate during the field ramp and a single characteristic relaxation time. It does not explicitly account for the detailed field dependence of $\Delta T_{\mathrm{ad}}$, thermal gradients within the sample, or field-dependent heat capacity. Despite these simplifications, the model captures the essential features of the experimental temperature evolution and provides a robust estimate of the adiabatic temperature change. Importantly, the extracted $\Delta T_{\mathrm{ad}}$ should be interpreted as an effective value under quasi-adiabatic conditions, with the true adiabatic limit expected to be slightly higher due to unavoidable thermal losses during the measurement.

\clearpage

\section*{Supplementary Note 5: Paramagnetic reference model for magnetic entropy}

To quantify the deviation of EuB$_6$ from conventional paramagnetic behavior, we construct a reference model based on non-interacting localized magnetic moments. This model provides a baseline for evaluating the magnetic entropy and its suppression under applied magnetic fields (as shown in Fig.~4a and b in the Main Text).

We consider a system of independent Eu$^{2+}$ ions, each carrying a total angular momentum $J = 7/2$ with Land\'e $g$-factor $g = 2$. In the presence of a magnetic field $H$, the energy levels are given by Zeeman splitting,
\begin{equation}
E_m = - g \mu_B m H, \quad m = -J, -J+1, \dots, J,
\end{equation}
where $\mu_B$ is the Bohr magneton. The partition function is
\begin{equation}
Z = \sum_{m=-J}^{J} \exp\left( \frac{g \mu_B m H}{k_B T} \right).
\end{equation}
The magnetic entropy of this simple paramagnet is then obtained from
\begin{equation}
S_{\mathrm{pm}}(T,H) = k_B \ln Z - \frac{1}{T} \left( \frac{\partial \ln Z}{\partial \beta} \right),
\end{equation}
where $\beta = 1/(k_B T)$. Equivalently, the entropy can be expressed as
\begin{equation}
S_{\mathrm{pm}}(T,H) = k_B \left( \ln Z - \beta \langle E \rangle \right),
\end{equation}
with $\langle E \rangle$ the thermal average of the energy.

In the absence of a magnetic field, all spin states are degenerate and the entropy reaches its maximum value,
\begin{equation}
S_{\mathrm{pm}}(T,0) = k_B \ln (2J+1).
\end{equation}
For Eu$^{2+}$ with $J = 7/2$, this gives
\begin{equation}
S_{\mathrm{pm}}(T,0) = k_B \ln 8.
\end{equation}
This value is plotted as the zero-field paramagnetic limit in Fig.~4a in the Main Text. When expressed per mole of Eu ions, the corresponding entropy is
\begin{equation}
S_{\mathrm{pm}} = R \ln 8,
\end{equation}
where $R$ is the gas constant.
Under an applied magnetic field, the degeneracy of the spin states is lifted, leading to a reduction in magnetic entropy as the spins become progressively aligned. In this model, the entropy reduction is solely determined by the ratio $g \mu_B H / k_B T$, and therefore exhibits a relatively gradual suppression with increasing field.

This paramagnetic reference provides a direct comparison for the experimentally extracted magnetic entropy of EuB$_6$. The experimentally determined magnetic entropy of EuB$_6$ shows a significantly stronger suppression under applied magnetic fields than predicted by the paramagnetic model, particularly in the temperature range near and above the Curie temperature. This deviation indicates that the magnetic degrees of freedom in EuB$_6$ are not independent, but instead exhibit strong correlations that enhance the field responsiveness. The failure of the non-interacting paramagnetic model to describe the observed entropy behavior highlights the importance of collective magnetic effects. In particular, the enhanced field-induced entropy reduction is consistent with the presence of magnetic clusters with large effective moments, which respond more strongly to external fields than individual spins. This provides additional support for the magnetic polaron picture discussed in the main text. We note that this reference model neglects exchange interactions, short-range correlations, and collective excitations. While it provides a useful baseline, it is not expected to capture the full complexity of the magnetic behavior in EuB$_6$, especially in the vicinity of the magnetic transition.
\clearpage

\section*{Supplementary Note 6: Langevin-type analysis of low-field magnetization}

To quantify the effective size of magnetic clusters in the paramagnetic regime, we analyze the low-field magnetization $M(H)$ using a phenomenological model consisting of a Langevin-type contribution from correlated magnetic clusters and a linear paramagnetic background from uncorrelated spins. This approach is analogous to treatments used in superparamagnetic systems, but here the extracted cluster size reflects emergent magnetic polarons rather than pre-existing nanoparticles~\cite{bansmann2005magnetic,sakamoto2016magnetization}. The total magnetization is expressed as
\begin{equation}
M(H,T) = M_{\mathrm{cl}}(H,T) + M_{\mathrm{lin}}(H,T),
\end{equation}
where $M_{\mathrm{cl}}$ represents the contribution from magnetic clusters and $M_{\mathrm{lin}}$ accounts for the linear paramagnetic background. The cluster contribution is modeled using a Langevin function:
\begin{equation}
M_{\mathrm{cl}}(H,T) = M_s(T)\,\mathcal{L}\!\left( \frac{\mu_{\mathrm{cl}}(T)\,H}{k_B T} \right),
\end{equation}
where $\mathcal{L}(x) = \coth(x) - 1/x$ is the Langevin function, $M_s(T)$ is the saturation magnetization associated with the cluster component, $\mu_{\mathrm{cl}}(T)$ is the effective magnetic moment of a cluster, $k_B$ is the Boltzmann constant, and $T$ is the temperature. The linear background is written as
\begin{equation}
M_{\mathrm{lin}}(H,T) = \chi_0(T)\,H,
\end{equation}
where $\chi_0(T)$ is an effective paramagnetic susceptibility that captures contributions not included in the cluster response. This includes the response of uncorrelated or weakly correlated Eu$^{2+}$ moments in the low-field limit, as well as possible additional background contributions such as Van Vleck and Pauli susceptibilities. The effective cluster moment $\mu_{\mathrm{cl}}$ is related to the number of Eu$^{2+}$ spins within a cluster:
\begin{equation}
\mu_{\mathrm{cl}}(T) = N_{\mathrm{cl}}(T)\,\mu_{\mathrm{Eu}},
\end{equation}
where $\mu_{\mathrm{Eu}} \approx 7\,\mu_B$ is the magnetic moment of a single Eu$^{2+}$ ion and $N_{\mathrm{cl}}(T)$ is the effective cluster size (number of Eu spins). The extracted $\mu_{\mathrm{cl}}(T)$ therefore provides a direct estimate of $N_{\mathrm{cl}}(T)$.

The fitting is performed on isothermal magnetization curves in the temperature range 12--35~K, focusing on the low-field region ($\mu_0 H \leq 0.3$~T). This field range is chosen to capture the nonlinear response associated with cluster alignment while avoiding high-field saturation effects not described by the simple Langevin model. For each temperature, the fitting parameters are:
$M_s(T)$: saturation magnetization of the cluster component, $\mu_{\mathrm{cl}}(T)$: effective cluster moment,
and $\chi_0(T)$: linear background susceptibility. The fitting is carried out using nonlinear least-squares regression. To improve stability, initial guesses for $\mu_{\mathrm{cl}}$ are chosen based on the curvature of $M(H)$ at low field, while $\chi_0$ is initialized from the high-field slope within the fitting window. Fitting results at selected temperatures are shown here in Fig.~\ref{fig:nonlinear} and fitted parameters as a function of temperature are summarized in Fig.~\ref{fig:langevin}.

For comparison, to assess whether the low-field magnetization can be described by independent local moments, we also analyzed the data using a conventional Brillouin-function model. In this picture, each Eu$^{2+}$ ion is treated as an isolated spin with total angular momentum $J = 7/2$ and Land\'e $g$-factor $g \approx 2$, yielding a fixed microscopic magnetic moment. The magnetization is expressed as
\begin{equation}
M(H,T) = A_B(T)\, B_J\!\left(\frac{gJ\mu_B H}{k_B T}\right) + B_B(T)\,H,
\end{equation}
where $B_J(x)$ is the Brillouin function. Here $A_B(T)$ is a temperature-dependent amplitude accounting for the total number of contributing moments (in units of emu/mol), and $B_B(T)$ captures a linear background susceptibility. Within this model, the field scale of the nonlinear response is fully determined by the single-ion moment and temperature, with no adjustable parameter controlling the curvature.

We applied this model to the same low-field regime ($H \leq 0.3$~T) and temperature range as used in the Langevin analysis. The comparison is also presented in Fig.~\ref{fig:nonlinear}. It is clearly seen that the Brillouin model systematically underestimates the pronounced curvature of $M(H)$, particularly in the vicinity of $T_c$. This discrepancy reflects the limited field responsiveness of independent Eu$^{2+}$ moments, whose characteristic energy scale $gJ\mu_B H$ remains small compared to $k_B T$ in this regime. The failure of the Brillouin description therefore indicates that the relevant field-responsive entities are not isolated spins, but more likely possess significantly larger effective moments, consistent with a cluster-based (magnetic polaron) picture.

\begin{figure}[!htb]
\includegraphics[width=0.7\textwidth]{SI_Figures/nonlinear_M.png}
\caption{\textbf{Fitting of low-field magnetization using a Langevin-type (solid line) or Brillouin-type (dashed line) model.} Experimental data are represented by open circles.
} 
\label{fig:nonlinear}
\end{figure}

\begin{figure}[!htb]
\includegraphics[width=\textwidth]{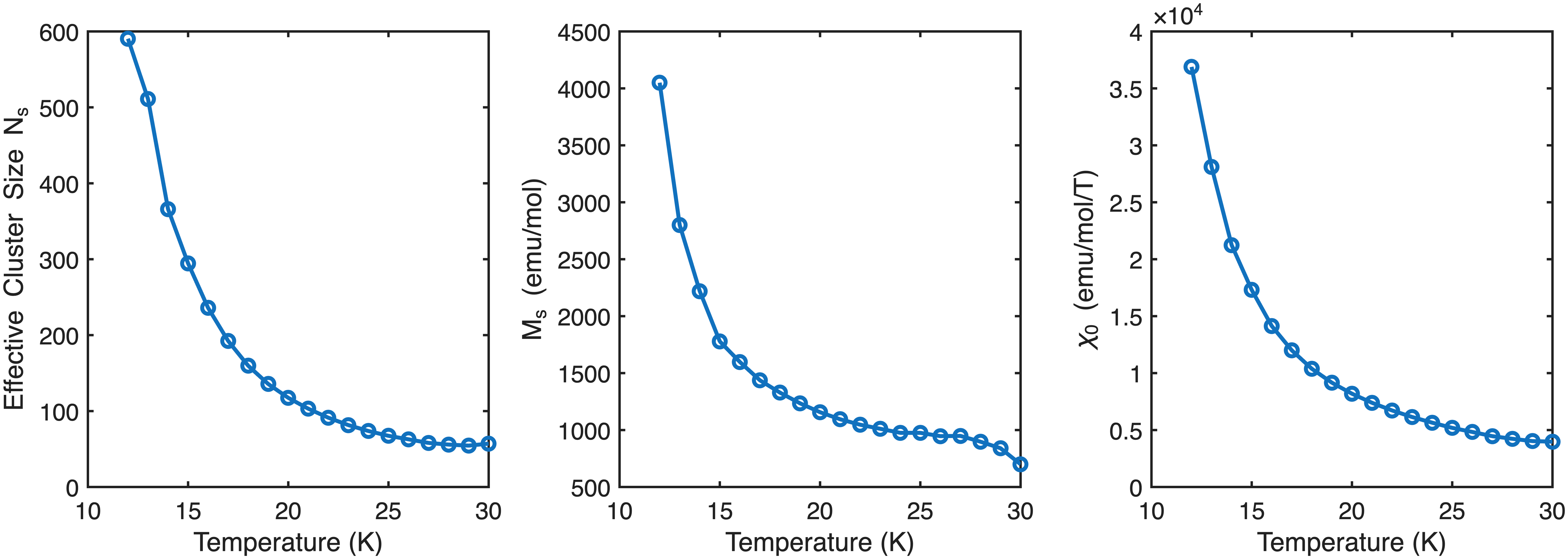}
\caption{\textbf{Fitted parameters using the Langevin model.} 
} 
\label{fig:langevin}
\end{figure}

The phenomenological Langevin model captures the coexistence of two magnetic components in the paramagnetic regime: (i) correlated clusters that exhibit nonlinear, saturating behavior and (ii) a background of weakly correlated spins that respond linearly to magnetic field. The presence of a significant Langevin component and its strong temperature dependence provide evidence for nanoscale magnetic clusters. The extracted cluster size $N_{\mathrm{cl}}(T)$ reaches values on the order of 300 to 400 spins near the Curie temperature (inset of Fig.~4d in the Main Text), corresponding to nanoscale regions with characteristic sizes of a few nanometers. The cluster size decreases with increasing temperature, consistent with thermal destabilization of magnetic polarons. We note that this model is phenomenological and does not explicitly account for interactions between clusters, size distributions, or dynamic fluctuations. At higher fields or far from the transition temperature, deviations from the model are expected. Nevertheless, the approach provides a robust and physically transparent estimate of the characteristic cluster size and its temperature dependence.
\clearpage

\section*{Supplementary Note 7: Large Negative Magnetoresistance in EuB$_6$}

\begin{figure}[!htb]
\includegraphics[width=\textwidth]{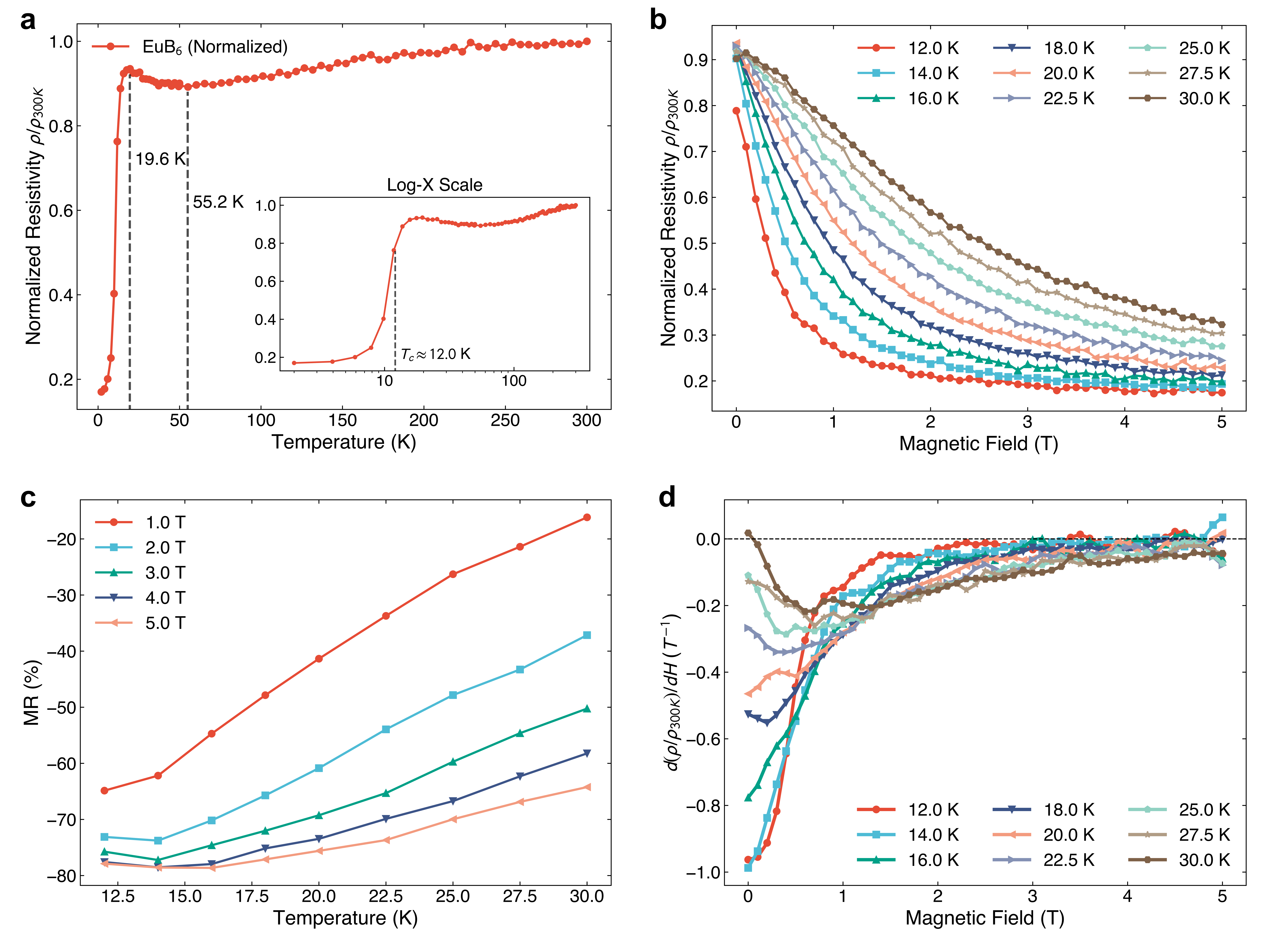}
\caption{\textbf{Magnetotransport signatures of magnetic polaron formation in EuB$_6$.}
\textbf{a,} Temperature dependence of the electrical resistivity from 300~K to 2~K. \textbf{Inset:} Resistivity plotted on a logarithmic temperature scale, highlighting the magnetic transition near $T_c \approx 12$~K.
\textbf{b,} Field-dependent resistivity $\rho(H)$ at selected temperatures, showing a large negative magnetoresistance, particularly in the vicinity of and above $T_c$.
\textbf{c,} Magnetoresistance (MR) ratio as a function of temperature for fields up to 5~T.
\textbf{d,} Differential magnetoresistance $d\rho/dH$ as a function of magnetic field. 
} 
\label{fig:MR}
\end{figure}

We conducted magnetotransport measurements of a polycrystalline EuB$_6$ sample that provide complementary information for the magnetic polaron regime (Fig.~\ref{fig:MR}) with good agreement with previous reports~\cite{sullow2000metallization,guy1980charge,zhang2009nonlinear}. The temperature-dependent resistivity from 300 K to 2 K (Fig.~\ref{fig:MR}a) displays three different regimes. Above 55 K, the resistivity shows typical metallic behavior and decreases with decreasing temperature. An upturn of resistivity is observed below $\sim 55$~K, signaling carrier localization associated with the formation of magnetic polarons, followed by a sharp decrease below $\sim 20$~K as these polarons grow, overlap, and coalesce into a percolating ferromagnetic state. Correspondingly, a very large negative magnetoresistance reaching -80\% under 5 T is observed (Fig.~\ref{fig:MR}b), with the strongest response occurring just above $T_c$. The magnitude of the magnetoresistance (Fig.~\ref{fig:MR}c) and the pronounced low-field peak in the differential response $d\rho/dH$ (Fig.~\ref{fig:MR}d) demonstrate that small magnetic fields can dramatically modify the transport properties, reflecting the field-induced alignment, growth, and stabilization of magnetic clusters.

\clearpage

\section*{Supplementary References}
\bibliography{references.bib}%